\documentclass{aastex}
\usepackage{emulateapj5}
\slugcomment{to appear in the ApJ Supplements}

\shorttitle{NEP survey: The Optical Identifications}
\shortauthors{Gioia et al.}

\def\deg{\hbox{$^\circ$}}
\def\sun{\hbox{$\odot$~}}
\def\arcmin{\hbox{$^\prime$}}

\def\hawaii{Hawai$'$i~}
\def\arcsec{\ifmmode^{\prime\prime}\;\else$^{\prime\prime}\;$\fi}

\newcommand{\myemail}{gioia@ira.cnr.it}

\begin{document}

\title{The {\em ROSAT\/} North Ecliptic Pole Survey: the  \\
Optical Identifications}

\author{I. M. Gioia\altaffilmark{1} \\ \myemail}

\affil{Istituto di Radioastronomia del CNR,
Via Gobetti 101, I-40129, Bologna, Italy \\
Institute for Astronomy, University of Hawai$'$i,
2680 Woodlawn  Drive, Honolulu, HI 96822, USA}

\author{J. P. Henry\altaffilmark{1}}
\affil{Institute for Astronomy, University of Hawai$'$i,
2680 Woodlawn  Drive, Honolulu, HI 96822, USA}

\author{C. R. Mullis\altaffilmark{1}}
\affil{European Southern Observatory, Karl-Schwarzschild-Str. 2,
Garching bei M\"unchen, D-85740, Germany}
\affil{Institute for Astronomy, University of Hawai$'$i,
2680 Woodlawn  Drive, Honolulu, HI 96822, USA}

\author{H. B\"ohringer, U. G. Briel, W. Voges}
\affil{Max-Planck-Institut fur Extraterrestrische Physik,
Giessenbachstrasse, Postfach 1312, Garching, D-85741 Germany}
\and

\author{J. P. Huchra\altaffilmark{2}}
\affil{Harvard-Smithsonian Center for Astrophysics, 60 Garden Street,
Cambridge, MA, 02138 USA}

\altaffiltext{1}{Visiting Astronomer at Canada-France-Hawai$'$i Telescope,
operated by the National Research Council of Canada, le Centre
National de la Recherche Scientifique de France and the University
of Hawai$'$i, and at the W. M. Keck Observatory, jointly
operated by the  California Institute of Technology, the
University of California and the National Aeronautics and Space
Administration.} 
\altaffiltext{2}{Observations reported here were made at the Multiple
Mirror Telescope Observatory, a joint facility of the Smithsonian Institution
and the University of Arizona.}

\begin{abstract}
The X-ray data around the North Ecliptic Pole (NEP) of the {\em ROSAT\/}
All Sky Survey  have been used to construct a contiguous area survey  
consisting of a sample of 445 individual X-ray sources
above a flux of $\sim$ 2$\times$10$^{-14}$ erg cm $^{-2}$ s$^{-1}$ in
the 0.5$-$2.0 keV energy band.
The NEP survey is centered at $\alpha_{2000}= 18^{h} 00^{m}, 
~\delta_{2000} =+66\deg 33\arcmin$ and covers a region of 80.7 deg$^{2}$
at a moderate Galactic latitude of {\em b\/} $=$ 29.8\deg.
Hence, the NEP survey is as deep and covers a comparable solid angle to
the {\em ROSAT\/} serendipitous surveys, but is also contiguous.
We have identified 99.6\% of the sources and determined redshifts
for the extragalactic objects.
In this paper we present the optical identifications  of the NEP catalog
of X-ray sources including basic X-ray data and  properties of the
sources. We also  describe with some detail the optical identification
procedure. The classification of the optical counterparts  to the NEP
sources is very similar to that of previous surveys,  in particular the
Einstein Extended Medium Sensitivity Survey (EMSS). The main constituents 
of the catalog are active galactic nuclei ($\sim$49\%), either type 1
or type 2 according to the broadness of their permitted emission lines.
Stellar counterparts are the second most common identification class
($\sim$34\%). Clusters and groups of galaxies comprise 14\%, and BL
Lacertae objects 2\%. One non-AGN galaxy, and one planetary nebula  
have also been found.  The NEP catalog of X-ray sources 
is a homogeneous sample of astronomical objects featuring  
complete optical identification. 

\end{abstract}

\section{Introduction}

Since their appearance in the 1970s, X-ray surveys  played the
major  role  in our understanding of the X-ray universe. The goals of 
these surveys are to provide a detailed accounting of the classes
of discrete sources which make up the X-ray sky, and also to define
large, complete samples of X-ray selected objects for statistical 
and individual  studies. Statistical analyses of flux-limited and 
optically identified  samples of astronomical objects supply important 
information on quantities such as source counts, luminosity functions, 
cosmological evolution and X-ray background contributions of the 
different populations.  Clusters of galaxies are a class of X-ray 
sources in which we are particularly interested as they are
important probes for the evolution of cosmic structure and for the
test of cosmological models \citep{gio01}. The advantage to construct 
cluster samples for cosmological studies from X-ray surveys
is the more direct relation between  luminosity and mass. The derived 
X-ray luminosity function is closely  related to the mass function 
of the clusters which is used as an important calibrator of the amplitude 
of the density fluctuation  power spectrum. 
The first X-ray imaging instruments onboard the {\em Einstein\/} Observatory 
allowed  the construction of the EMSS  catalogs  of active galactic 
nuclei  (AGN), galaxy clusters, BL Lac objects  and stars  \citep{gio90a, 
sto91, mac94}  and the discovery of the negative evolution of the space
density of X-ray clusters  \citep{gio90b, hen92}. 
Given the diverse  nature of the X-ray emission  from the various X-ray 
counterparts (e.g. diffuse hot cluster gas,  non-thermal AGN emission, 
stellar coronae) a variety of complementary  observations were required to 
correctly identify the optical  objects associated with the EMSS  X-ray 
sources. The EMSS was constructed using those objects found 
serendipitously in each field  after excluding the central region of 
the field including the target itself.

{\em ROSAT\/} \citep{tru83} was  the first X-ray imaging experiment  
to survey the  entire sky. After  its launch a greatly increased  
sensitivity and source location accuracy over all previous X-ray sky 
surveys  were available.  Many X-ray  surveys were compiled either from  
the {\em ROSAT\/} All-Sky-Survey data ({\em RASS\/};  
\citealp{vog99})  or from pointed data (i.e. EMSS-like surveys but 
mostly to search for clusters of galaxies). In general, the 
surveys extracted from the {\em RASS\/}, the contiguous area  surveys, 
cover  a very large solid  angle  ($\sim$ 10,000 deg$^{2}$ or more)
but are shallower  than the pointed data surveys. 
Given the large solid angle sampled and the sizes of the superstructures 
(several hundreds of Mpc), the contiguous area surveys, and in particular 
the all sky surveys,  are good tracers of large-scale structure. They allow 
investigations of the clustering properties of clusters and of the power
spectrum of their distribution.
The contiguous area surveys typically sample the nearby universe 
($z<0.3$) and are  used as an excellent local reference for cluster studies 
at higher redshift.   Among the contiguous area surveys  we mention the 
BCS \citep{eb98}, and its extension  eBCS \citep{eb00},  
RASS1-BS \citep{deg99}, NORAS \citep{boe00},   MACS
\citep{eb01}, REFLEX \citep{boe01} and the NEP survey  
\citep{mul01a, hen01, vog01}.

The great advantage of the serendipitous surveys, those extracted from 
the pointed data, is their much higher 
sensitivity, about two orders of magnitude deeper than the contiguous 
area surveys ($\sim 10^{-14}$ vs $\sim 10^{-12}$ erg 
cm$^{-2}$ s$^{-1}$ in the {\em ROSAT\/} band), even though 
their solid angle is less  than $\sim$200 deg$^{2}$. 
The serendipitous surveys are not restricted  to 
the local universe, but probe z $>$ 4 for quasars and z $>$ 1 for 
clusters of galaxies. As with the contiguous area  surveys, many sources 
are common to the serendipitous surveys  since all use the same pointed 
{\em ROSAT\/}  data (see Table ~5 in \citealp{mul03}, for common 
cluster sources in all the present X-ray surveys). 
Among the serendipitous X-ray surveys  for clusters 
of galaxies we mention the  RDCS \citep{ros95, ros98}, RIXOS 
\citep{cas95, mas00}, Southern SHARC \citep{bur97, col97} and  Bright 
SHARC \citep{rom00}, 160 deg$^{2}$ \citep{vik98, mul03}, WARPS 
(\citealt{per02}, and  references therein), BMW \citep{cam99, laz01} and  
ROXS \citep{don02}.

We have used  data around the North Ecliptic Pole of the {\em RASS\/} to 
construct a contiguous area survey consisting of a homogeneous sample of 
445 X-ray  sources. The region around the NEP possesses the deepest 
exposure and consequently the greatest sensitivity of the entire 
{\em RASS\/}.  Hence,  the 9\deg$\times$ 9\deg ~survey region centered at 
$\alpha_{2000}= 18^{h} 00^{m},~ \delta_{2000} =+66\deg 33\arcmin$
covers the deepest, wide-angle  contiguous region ever observed in X-rays. 
This unique combination of  depth plus wide, contiguous solid angle provides 
the capabilities of detecting both high-redshift objects and large scale 
structure, which  were the aims of our involvement in the survey.

A comprehensive description of the {\em ROSAT\/} NEP Survey and the 
principal results are presented in \citet{mul01a}.\footnote{A link 
to his thesis can be  found at  
{\em http://www.ifa.hawaii.edu/$\sim$mullis/nep-phd.html\/}}
An overview of the NEP survey, including the selection function,
the optical identification program and number-count  distributions 
for the various classes of objects,  has been published in 
\citet{hen01}.  \citet{vog01} gives a summary of the X-ray data and the 
statistical properties of the NEP sources. \citet{gio01} present evidence  
for  cluster X-ray luminosity  evolution and \citet{mul01b} describe the 
details on the NEP supercluster found  in the survey X-ray data. 
In this paper we discuss in more detail the optical identification program,
the  methodology used to identify the sources, and we  present the optical 
catalog with a description of the optical properties of the sources. For 
convenience, we give the basic X-ray properties of the sources, but a more 
detailed description  of the X-ray data,  the detection algorithm and the 
statistical properties of the NEP sources will appear in a subsequent  
paper.   For consistency with previous work, we assume throughout the paper
a Hubbble constant of $H_{0} = 50 ~h_{50}$ km s$^{-1}$ Mpc$^{-1}$, a matter 
density parameter $\Omega_{M0} = 1$, and a cosmological constant of 
$\Omega_{\Lambda0}=0$.

\section{The X-ray data} \label{X-ray}

The NEP region was observed many times by the {\em ROSAT\/} satellite 
since the {\em RASS\/} scan pattern overlapped at the ecliptic poles. 
While the mean {\em RASS\/} exposure time across the entire sky 
is approximately 400 s, the NEP region exposure time approaches 40 ks at 
the pole. The minimum,  median and maximum exposure times in the NEP 
survey regions are 1.7, 4.8, and 38 ks, respectively. The 80.7 deg$^{2}$  NEP 
region is a good area to pursue unbiased  surveys of the extragalactic 
sky  since it is at a moderate Galactic latitude of  {\em b\/} $=$ 
29.8\deg ({\em l\/}  $=$ 96.4\deg) and has a mean neutral  column density of  
{\em $<n_{H}>$\/} $\approx$ 4.3$\times10^{20}$ cm$^{-2}$ \citep{elv94}.
The other pole of the ecliptic where the exposure is piling up, the South 
Ecliptic Pole region, or SEP, has much less exposure since the PSPC was
automatically shut off due to the enhanced charged particle density of 
the South Atlantic Anomalies which could damage the detector. Furthermore
the SEP is in the vicinity  of the Large Magellanic Clouds, a crowded 
stellar region which is not amenable for extragalactic work. All these
considerations  make the NEP survey  unique.

The data used in our work were extracted from the second processing
of the  {\em ROSAT\/} data  ({\em RASS-II\/}) which has 
improved attitude quality, improved spline-fitted background map 
and fully merged photon data \citep{vog99}. 
The detection  algorithm  of the {\em RASS\/} is based on a multi-scale, 
sliding detect aperture \citep{vog99}.  Candidate sources identified by 
this procedure,  operating at a low  acceptance threshold, were passed  
to a  maximum-likelihood (ML) algorithm for more accurate determinations of 
source existence likelihood and other interesting parameters
\citep{vog01, mul01a}. 

To be part of the NEP survey an X-ray source must be in the 
right ascension  range $17^{h} 15^{m} < \alpha_{2000} < 18^{h} 45^{m}$, 
and in the declination range  $62\deg < \delta_{2000} < 71\deg$;
its maximum likelihood of existence must be $L\geq$10 ~where
L $= -$ln P, and  P is  the probability that the source count  rate is 
zero; its signal-to-noise ratio for source count rate must be greater than 
4$\sigma$. The detection energy band  used  is the {\em ROSAT\/} broad 
band  0.1$-$2.4 keV. There are 445 unique sources within the survey region 
fulfilling these requirements. Twenty-one multi-detections of individual, 
often extended, sources were removed in constructing the final sample.
Approximately 2 of the 445 sources are expected to be spurious considering 
the survey solid angle and telescope beam size appropriate for the
{\em RASS\/} (half power diameter $=$ 3.1\arcmin, \citealp{boese00}).
Of the 515 additional X-ray sources with a likelihood of
existence $L<$10 in the NEP region, only 3 have a signal-to-noise 
ratio of  $>4\sigma$. Hence, the principal selection criterion for 
the NEP survey is  the 4$\sigma$ threshold in count rate significance.

The observed count rate for each source is extracted from a 5\arcmin ~radius 
circular aperture (except for one source, RX\,J1834.1+7057, where 
r$_{extract}=$6.5\arcmin). Different model spectra, according to the source  
nature,  have been  adopted to convert from count rates to detect flux. 
An additional correction is applied to the detect flux of
extragalactic X-ray sources to account for the effect of Galactic  
absorption along the line of sight. Unabsorbed fluxes were derived
using neutral hydrogen ($n_{H})$ column density data
obtained from the 21 cm observations of  \citet{elv94},  supplemented by 
\citet{sgw92}, at the location of  the X-ray source.  The minimum, median, 
and maximum neutral $n_{H}$ for the NEP sources are 
(2.4, 4.2, 8.2)$\times10^{20}$ cm$^{-2}$, respectively.

For  AGN  we have used a power-law spectrum 
with an energy index of $-$1, resulting in a median conversion factor 
from  count rate (0.1$-$2.4 keV) to flux (0.5$-$2.0 keV) of 
$1.20\times10^{-11}$, with a $\pm12\%$ variation over the full range of column
densities sampled by the NEP AGN. For galaxy clusters a  Raymond-Smith 
plasma spectrum was adopted, with a metallicity of 0.3 solar plus a plasma 
temperature, and a measured redshift particular to  each cluster. 
The gas temperature of the intracluster medium is estimated using
the $L_{X}-kT$ relation of \citet{whi97}, $kT=$~2.76~keV 
L$^{0.33}_{X,bol,44}$,
where $ L^{0.33}_{X,bol,44}$ is the bolometric X-ray luminosity in units 
of $10^{44}$ erg s$^{-1}$.  An iterative process begins by assuming a 
temperature of 5 keV. Total flux is converted to a K-corrected 
luminosity\footnote{Luminosities in the ROSAT rest-frame are transformed
in luminosities into the object's rest frame using the K-correction, in the 
$0.5-2.0$ keV band, defined as $k_{0.5-2.0} = {{\int^{2.0}_{0.5} f_{E}dE} \over
{ \int^{2.0 (1+z)}_{0.5 (1+z)} f_{E}dE}}$ where $f_{E}$ is the differential
flux (flux per unit energy) as a function of energy and the integration 
limits are energy band edges in keV.}
in the $0.5-2.0$ keV band,  which is converted to a 
bolometric luminosity. This preliminary $L_{X,bol,44}$ is  used to predict 
the cluster  temperature from the $L_{X}-kT$ relation above. The procedure is 
repeated until the estimated temperature converges. For clusters of galaxies 
the flux  conversion factor has a median of $1.07\times10^{-11}$ and 
varies by $\pm16\%$ over the full range of column  densities 
and redshifts sampled by the NEP clusters. For stars a Raymond-Smith 
plasma spectrum with a temperature of $10^{7}$K, a solar metallicity 
abundance, and no  Galactic absorption were assumed, resulting in a 
conversion factor  of $5.95\times10^{-12}$. 

To correct for flux outside the detect cell a size correction for point sources
has been applied by calculating the integral of the {\em RASS\/} point-spread 
function, or PSF \citep{boese00}, at 1 keV within the  circular detect 
aperture of $5\arcmin$ radius.
The correction applied is 1.0369 which corresponds to the 
reciprocal of the integral of the PSF within the circle. For extended 
sources like galaxy  clusters we adopted the $\beta$-model 
\citep{cff76} surface brightness distribution 
$I(r) = I_{o}[1 + (r/r_{c})^{2}]^{-(3\beta-0.5)}$, with a core radius, 
r$_{c}$, equal to 0.25 Mpc with $\beta=2/3$. The $\beta$ model surface 
brightness  is  convolved with the  {\em RASS\/}  PSF and integrated out to 
infinity\footnote{The difference between integrating out to infinity
and to 12 core radii is about 8\%}.  Then this integral is divided 
by the integral within the 5\arcmin ~circle  aperture to obtain the size 
correction. The median size correction for clusters is 1.33. At the median 
redshift of the NEP  cluster sample ($z=$0.2)  the size  correction varies 
by only $\pm5\%$ for core radii between 0.2 and 0.3 Mpc. 

K-corrected luminosities are computed for the extragalactic objects in
the  NEP survey using the redshift particular to each source. 
For clusters of galaxies the K-corrections were computed 
assuming again a Raymond-Smith plasma spectrum with a metallicity of 
0.3 solar. The median K-correction for the NEP clusters is 0.95. For 
AGN the adopted  power-law spectrum with energy index $-$1 results in a
K-correction equal to 1. 
Unabsorbed detect fluxes (f$_{det}$) and total fluxes (f$_{tot}$) in 
the {\em ROSAT\/} rest frame are reported  in Table~3   
(see \S\,\ref{Catalog}),  together  with K-corrected luminosities in 
the source rest frame (all in the {\em ROSAT\/} hard  band 0.5$-$2.0 keV). 
 
The selection function for the NEP sources is described in detail 
by \citet{mul01a} and \citet{hen01}, including the {\em ROSAT\/} NEP 
sky  coverages for AGN, clusters and stars.

\section{Optical Identification Procedure} \label{Identification}

A comprehensive program of optical follow-up observations to determine
the nature of each of the X-ray sources in the NEP sample led
to the identification of  all but two NEP sources. There is evidence 
that at least one of the two unidentified sources is a blend and that 
both may be statistical fluctuations. Note, two spurious sources are 
expected in a sample of 445  objects (see  \S\,\ref{X-ray}).
The NEP is observable from mid-May until mid-August from 
Hawai$'$i. The optical identification program began in the summer of 1991,  
just a few months after the {\em RASS\/} was finished, and 
ended in the summer  
of 2000. We made most of the optical observations from Mauna Kea where 126 
nights were assigned of which 100 were clear. We spent 101 nights at the 
University of \hawaii (UH) 2.2m telescope, 17 nights at the 
Canada-France-Hawai$'$i (CFH) 3.6m telescope and 8 nights at the Keck 10m 
telescope. Additional observations  were made at Mt Hopkins where 5 nights at 
the Multiple Mirror Telescope (MMT) were  assigned, of which 2 were clear and 
a clear night of 1.5m telescope was  also used. In total we used 132 
nights of which 103 were clear (see Table~\ref{table1}).

\begin{figure}[t]
\epsscale{1}
\plotone{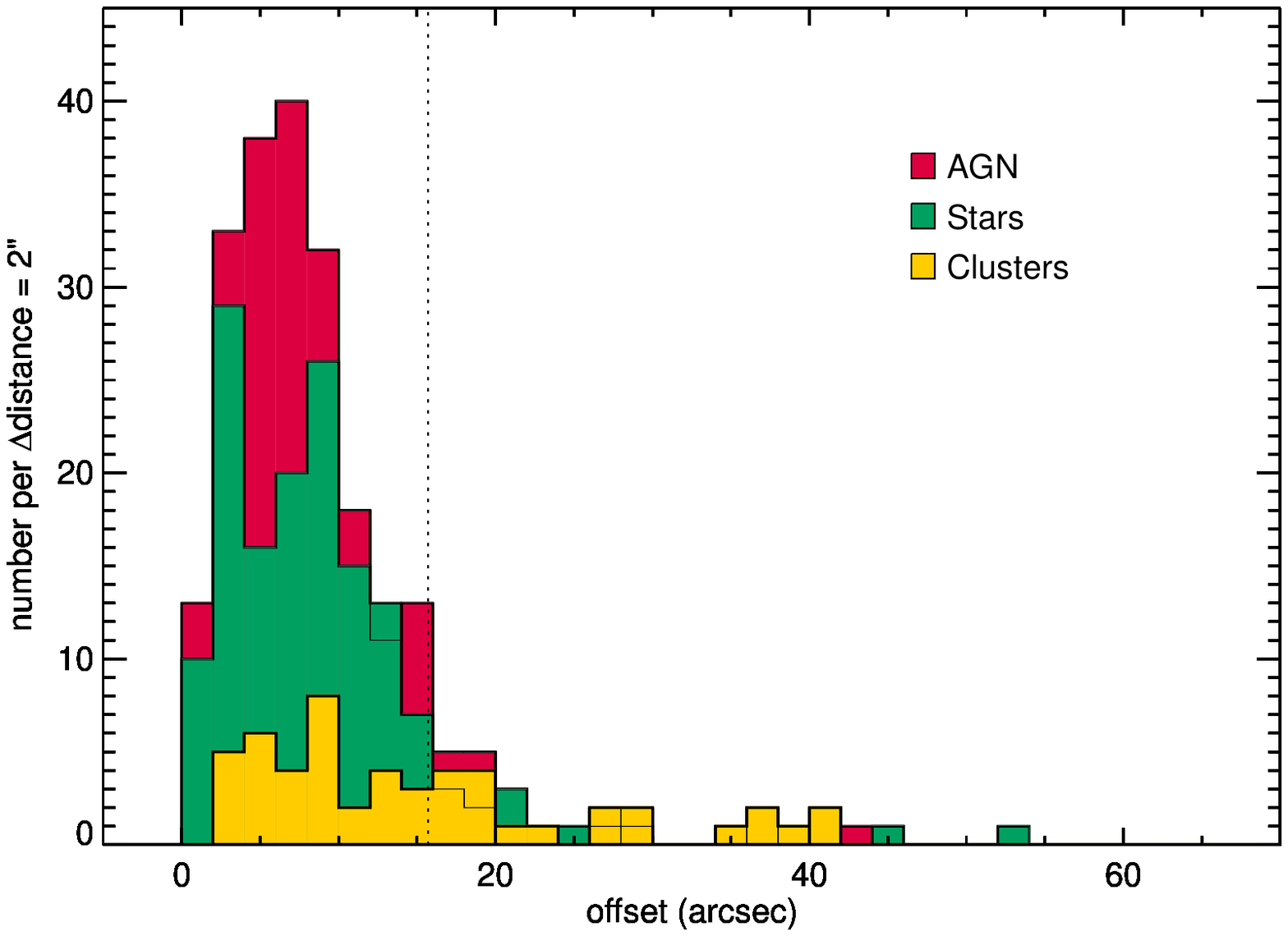}
\caption[ROSAT NEP point source offsets between optical and X-ray positions]
{Angular position offsets between the NEP X-ray sources 
and their optical counterparts.
AGN are shown in dark grey, stars in medium grey and galaxy clusters in 
light grey.  The vertical line at 15.7\arcsec indicates the offset from 
the X-ray position  within which 90\% of the AGN and stars fall (adapted from 
Fig. 3.5 of Mullis 2001). }
\label{x-offset}
\end{figure}

The procedure used to identify an X-ray source is essentially that described 
in \citet{sto91} for the EMSS  survey. The process with the NEP 
survey is considerably easier since the positional uncertainty of 
the {\em RASS\/} is 15.7$''$, about 15 times smaller in area than the 
Einstein IPC (15.7\arcsec  vs. $\sim$ 60$''$).  The EMSS had between 1 and 8 
objects in its error circle visible on the POSS plates and often 
had to use an X-ray to optical flux criterion to discriminate between more 
than one plausible counterpart in their large error circle.  Whereas, 
for 93\% of NEP sources there are $\leq2$ objects in the error circle 
visible in the DSS and it  is very rare to have more than one plausible 
identification fall within the error circle. 
The 15.7\arcsec positional accuracy of the NEP survey has been computed by 
using the offset from the X-ray position within  which 90\% of the point 
sources (AGN and stars) fall. Fig.~\ref{x-offset}  shows an histogram of the
angular offsets between the X-ray sources and their optical counterparts
with the 15.7\arcsec offset indicated by the vertical dotted line.
The basic procedure is to spectroscopically examine objects in close proximity
on the sky to the X-ray source until a likely optical counterpart is located.
The search begins with the object closest to the X-ray source position or with
a blue stellar object if such a source is present in the positional error 
circle.

We started by inspecting finding charts of the 445  NEP sources 
prepared from the Automated Plate Machine (APM) object catalog
\citep{hook96}. The APM scanned both colors of the Palomar Observatory
Sky Survey photographic plates (POSS-I, \citealt{wil52}; California, 
Institute of Technology 1954) providing a matched object catalog down to 
$m = 21.5$ in blue ($O$, 3200\AA--4900\AA) and $m = 20$ in red ($E$, 
6200\AA--6800\AA). These passbands resemble the Johnson $B$ and 
Kron-Cousins $R$ except for the narrowness of the E passband. Magnitudes 
have an internal accuracy of 0.1 magnitude for objects brighter than 1 
magnitude above the plate limit, and a typical external accuracy of 0.5
magnitudes. The APM also classifies objects as either stellar, non-stellar, 
or a blend of multiple sources. Overlays of optical and X-ray images of the
NEP sources were also prepared using the second Palomar Observatory Sky 
Survey (POSS-II) and the {\em RASS\/} data in order to check the  APM 
classifications, resolve the blends, and extend the magnitude limit.  
Digitized  Sky  Survey (DSS) images of the POSS-II red plates were obtained
using the STScI {\em WWW\/} interface \footnote{
{\em http://archive.stsci.edu/dss\/}}. The full catalog of the  
finding charts and the overlaid X-ray  contours will 
appear in a forthcoming  paper (see also 
Appendix C in \citealt{mul01a}).  Through the inspection of the optical plus 
X-ray contour  finders, and aided  by the APM object  catalog and by 
cross-correlations with  SIMBAD and NED 
databases\footnote{{\em http://simbad.harvard.edu} and
{\em http://nedwww.ipac.caltech.edu}}, small  percentages of X-ray sources 
were immediately identifiable with previously known X-ray emitters (5\%) and 
very bright stars ($\sim$11\%). The bright star identifications were only made
where evidence for  alternate counterpart was absent. During the inspection of 
the  finders a preliminary  target list was compiled for spectroscopic 
follow-up of the objects  
within the 90\% confidence error circle of 15.7\arcsec. 
Ten percent of  the sources had 0 visible sources, 55\% had 1, 
28\% had 2, 5\% had 3 and 2\% had 4 or more visible sources within the 
error circle. Extremely blue stellar objects ($O-E\leq1.3$) were given 
priority as these sources are almost always AGN. Fields with an overdensity 
of  galaxies were flagged as potential cluster candidates as were fields that 
contain a few galaxies at the limits of the DSS finders.

The surface density of plausible X-ray counterparts and the positional 
uncertainties of the X-ray data are sufficiently low that only one plausible 
counterpart is likely to fall in the positional error circle. Hence, confusion
levels are negligible in the NEP survey and can be quantified. We have 
computed the number of objects that may be found in the error circle by chance.
Using the AGN optical surface densities from \citet{boy88}, a contamination 
rate of false-positive less than 1\% at  B$<$20.7\footnote{We have 
approximated the {\em B\/} magnitudes using the {\em O\/}  
magnitudes from the APM} 
is  obtained, only one ``random'' AGN is expected out of 190 AGN  
observed at  {\em B\/}$<$20.7. Using the Galactic model of \citet{wai92} to 
estimate the number of plausible stellar counterparts randomly present in 
the survey, we proceeded as follows.  Given the high certainty of the AGN 
identifications, we removed those sources leaving 228 NEP error circles 
which subtend 0.0136  deg$^{2}$ of the sky. At  $B < 17.2$, 137
F--M stars are observed where 12 coincidences with F--M are
expected. Thus the false-positive rate for $B < 17.2$  is 9\% (or 12/137).

Though detailed spectral classifications for the NEP stellar 
identifications was not attempted in all cases, it is possible to
reliably confirm M dwarfs due to the distinctive TiO absorption bands in
their spectra.  Approximately 16\% of the NEP X-ray sources identified
with stars are M dwarfs, of which 20 with $B < 18.6$. Using the solid angle 
of all non-AGN NEP sources (0.0136  deg$^{2}$) and considering the source 
density of M stars down to $B < 18.6$, only one M star would fall into the 
error circle by random chance. Consequently the false-positive rate for M 
dwarfs  down to $B < 18.6$ is 5\% (1/20) while the rate for the remaining 
F--K dwarfs is 6\% down to $B < 16.3$ (7 coincidences expected against 114 
observed). Hence, the random probability of a  late-type dwarf falling in the 
NEP  positional error circles is quite low, though not as low as for AGN. 

We refer the reader to Mullis (2001, see his Chapter 4) for 
additional details on  the optical  identification procedure.

\subsection{Optical Imaging Observations}   \label{Imaging}

A program of optical imaging was undertaken to observe those fields around 
the NEP  sources which exhibited no objects in the X-ray 
error circle of the  APM or DSS data. The distant cluster candidates 
were also observed. Images  in two colors, B and I (Kron-Cousins) bands, 
were obtained at the UH\,2.2m 
to discriminate between AGN (usually blue) and  the cores of galaxy 
clusters, predominantly populated by red, early-type galaxies.
We used the TEK\,2048$^{2}$ CCD at  the f/10 focus of the UH\,2.2m telescope 
which gives a scale of 0.22\arcsec pix$^{-1}$ and a field of view (FOV) of 
approximately $7.5 \times 7.5$ arcmin$^{2}$. The exposures were typically 
10 minutes in both bands resulting in limiting magnitudes of
$\sim23$ in $B$ and $\sim22$ in $I$. Additional 30 min exposures were 
obtained for sources that were eventually identified as distant clusters.
Photometric calibration of the imaging data used the M92 standards
\citep{chri85, hea86}.

\subsection{Spectroscopic Observations} \label{Spectroscopy}

In parallel to imaging we acquired low-resolution spectra of objects 
starting with the one closest to the X-ray source  position  and 
then going out to larger offsets
until the first plausible counterpart was found. Alternatively, if 
a blue stellar object was present in the positional error circle, the search
began with that object and then proceeded to other objects if necessary. In 
either case, if there was evidence of an overdensity of  faint objects close
to the X-ray source in the imaging data, additional spectra were obtained to 
test for the presence of a galaxy cluster. Particular attention was given to 
assure that clusters were not missed due to close projections with bright 
stars or due to AGN in the cluster environment.  
Most of the spectroscopy at the beginning was done using the UH\,2.2m
Wide-Field Grism Spectrograph (WFGS),  but  we also used the 
Multi-Object-Spectrograph (MOS) on the CFHT 3.6m, and the Low-Resolution 
Imaging Spectrograph (LRIS, \citealt{oke95}) on Keck.

At the UH\,2.2m telescope we used the  420 l mm$^{-1}$ red grism and a 
200$\micron$ slit (1.8 arcsec$^{-1}$) which provided at the f/10 focus a 
pixel  scale in spectroscopic mode of ~3.6 \AA ~pix$^{-1}$, a spectral 
resolution of $\sim19$ \AA ~FWHM, and a wavelength coverage of 
approximately 3800\AA--9000\AA. In imaging mode the FOV is 
$4.75\times4.75$ arcmin$^{2}$ and the image scale is 0.35\arcsec pix$^{-1}$.
Observations at the MMT were made with the MMT spectrograph using a
280$\micron$ slit (1 arcsec$^{-1}$) with the 300 l mm$^{-1}$ 
grating which provides a dispersion of 1.96 \AA ~pix$^{-1}$ and a spectral 
resolution of $\sim$6 \AA~ FWHM \citep{fs94}. Observations at the FLWO 
(F. L. Whipple Observatory) 1.5 m telescope were made with the photon 
counting  Reticon system \citep{lath82} and 3\arcsec slit which also 
provides a  resolution of $\sim$6 \AA ~FWHM.

The instrument setups  for other telescopes, although different, provided
similar performance. The  UH\,2.2m telescope was used to observe both stars 
and AGN, but also to acquire long-slit spectra of relatively
bright cluster galaxies. Multi-object spectra were subsequently taken 
with CFH and Keck telescopes for the fields around X-ray sources which
showed overdensities of faint galaxies. This multi-object spectroscopy
approach was a departure from the \cite{sto91} procedure and provides
more confidence in any cluster identifications since many concordant
redshifts could be obtained. In total over one thousand objects
were observed spectroscopically. Calibration data were based on 
spectrophotometric standard stars of \cite{oke83} and \cite{mas88}.

The optical data were analyzed using standard IRAF\footnote{{\em 
http://iraf.noao.edu\/}} reduction packages and IDL\footnote{{\em
http://www.rsinc.com\/}}  routines.  A quick reduction and
interpretation of long-slit spectra were normally completed in
``real-time'' at the telescope in order to decide if further spectra
were required to make a reliable identification of a particular 
X-ray source. Nonetheless, all data were subsequently reduced off-line 
using standard procedures.
Spectra were classified according to the presence or absence of various
absorption and emission lines and the shape of the continuum emission.
Once features have been identified, the redshift of a source was
measured based on the offset of the features from their restframe
wavelengths (typical redshift uncertainty $\delta z \la 0.001$).  
Commonly observed absorption lines included: the  Ca\,II H \& K doublet, 
the 4000\AA ~break, the G band, Mg\,Ib, Na\,Id, and H$\alpha$.  Often
present emission lines included: MgII, [O\,II]\,3727, H$\beta$, 
[O\,III]\,4959,  [O\,III]\,5007 and sometimes  H$\alpha$. 

\section{Identification Content} \label{Idstat}

\begin{figure}[t]
\epsscale{1}
\plotone{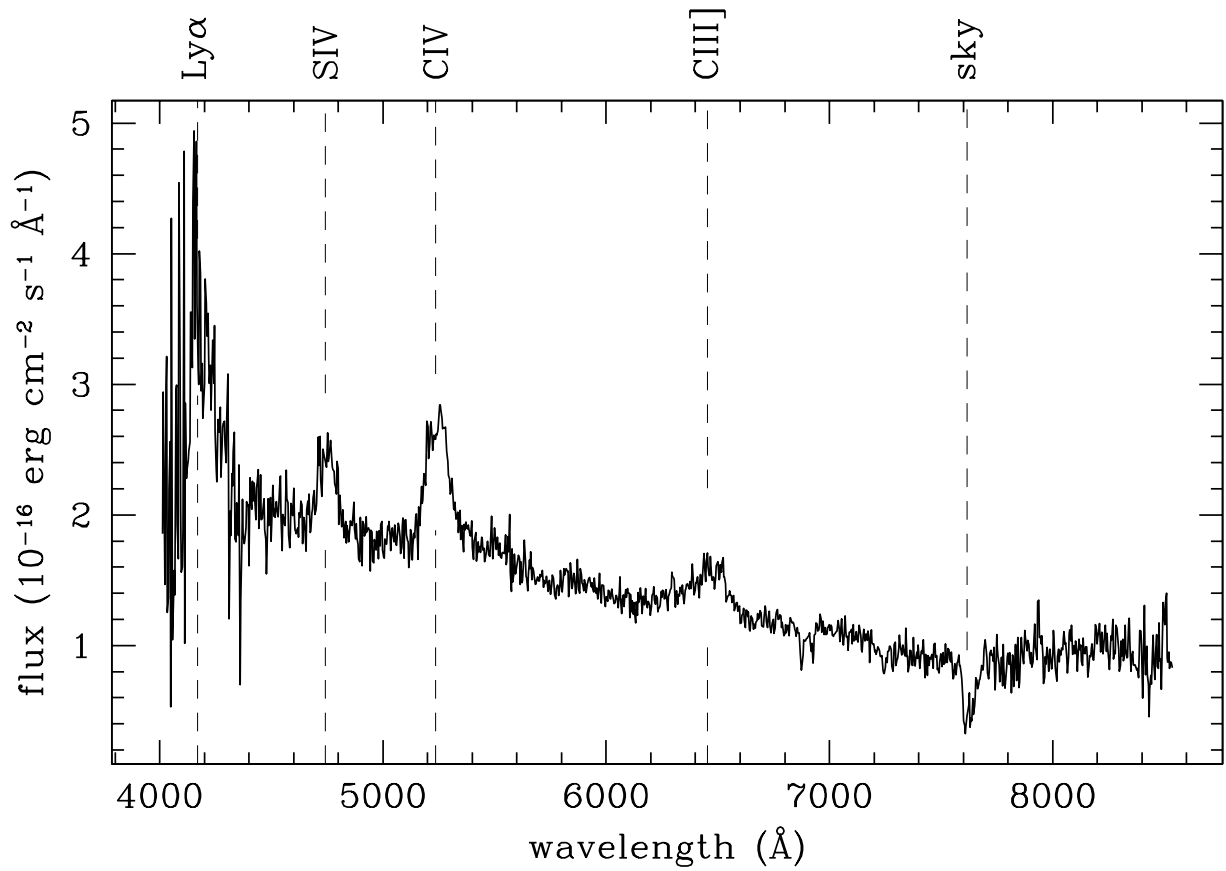}
\caption[RX\,J1747.1$+$6813 spectrum]
{RX\,J1747.1+6813: AGN1 at z$=$2.392$\pm$0.002 ($\alpha_{2000}=$~17~47~12.7, 
$\delta_{2000}=+$68~13~26). Longslit spectrum of  a distant QSO obtained 
with the UH\,2.2m telescope on 19 June 1998. The total integration 
time was 20 minutes. The   dashed lines indicate the positions of the 
emission lines at the AGN1 redshift.  Wavelengths of atmospheric 
absorption bands are  also indicated.}
\label{2810}
\end{figure}

\begin{figure}[t]
\epsscale{1}
\plotone{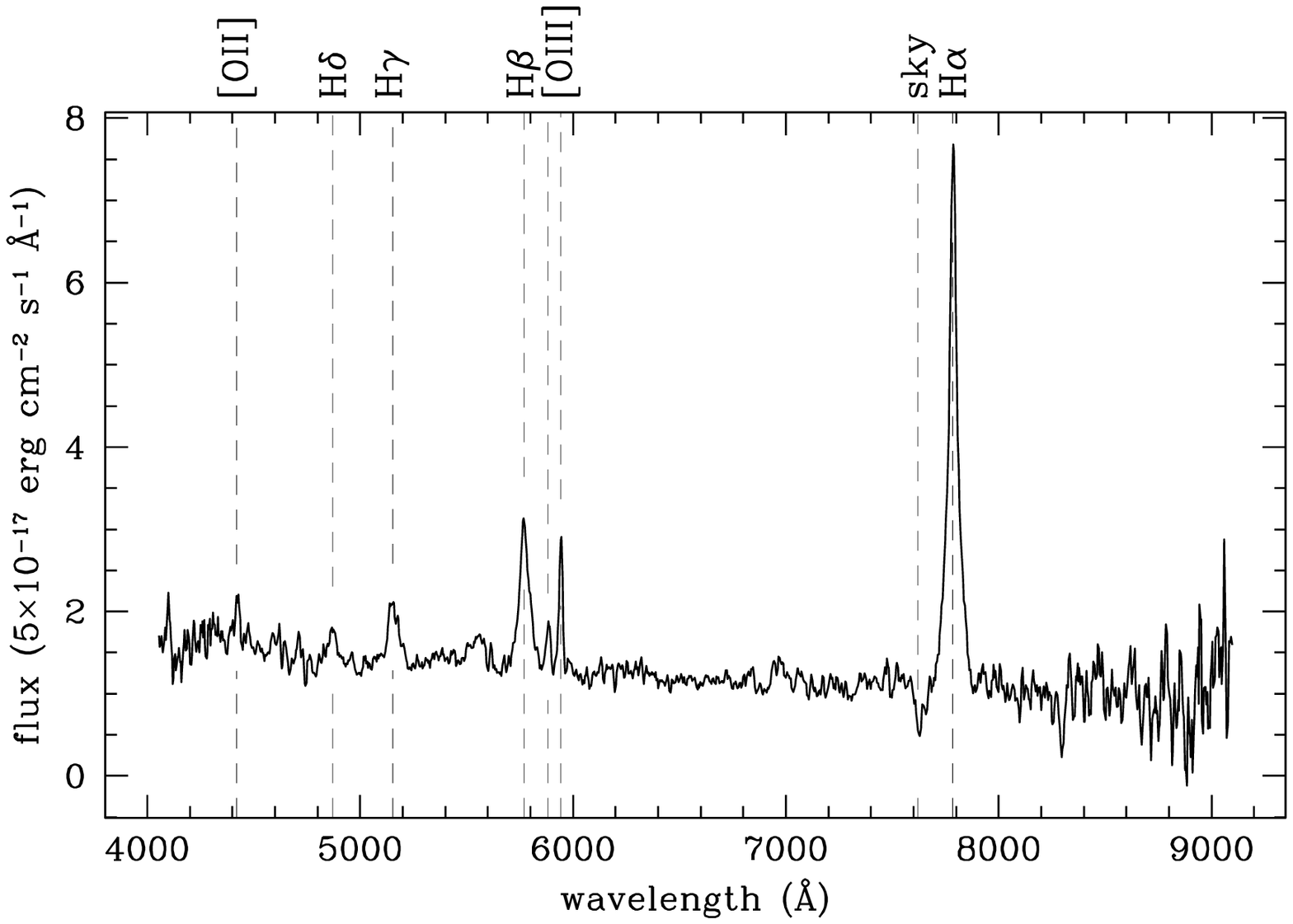}
\caption[RX\,J1758.0$+$6851 spectrum]
{RX\,J1758.0$+$6851: AGN1 at z$=$0.1876$\pm$0.0005 
($\alpha_{2000}=$~17~58~03.7, $\delta_{2000}=+$68~51~51). 
Longslit spectrum of a Sy1 galaxy obtained 
with the UH\,2.2m telescope on 5 June 1994. The total integration time was 
30 minutes. The dashed lines indicate the positions of the emission lines 
at the AGN1  redshift.  Wavelengths  of atmospheric absorption bands are  
also indicated.}
\label{3620}
\end{figure}

\begin{figure}[t]
\epsscale{1}
\plotone{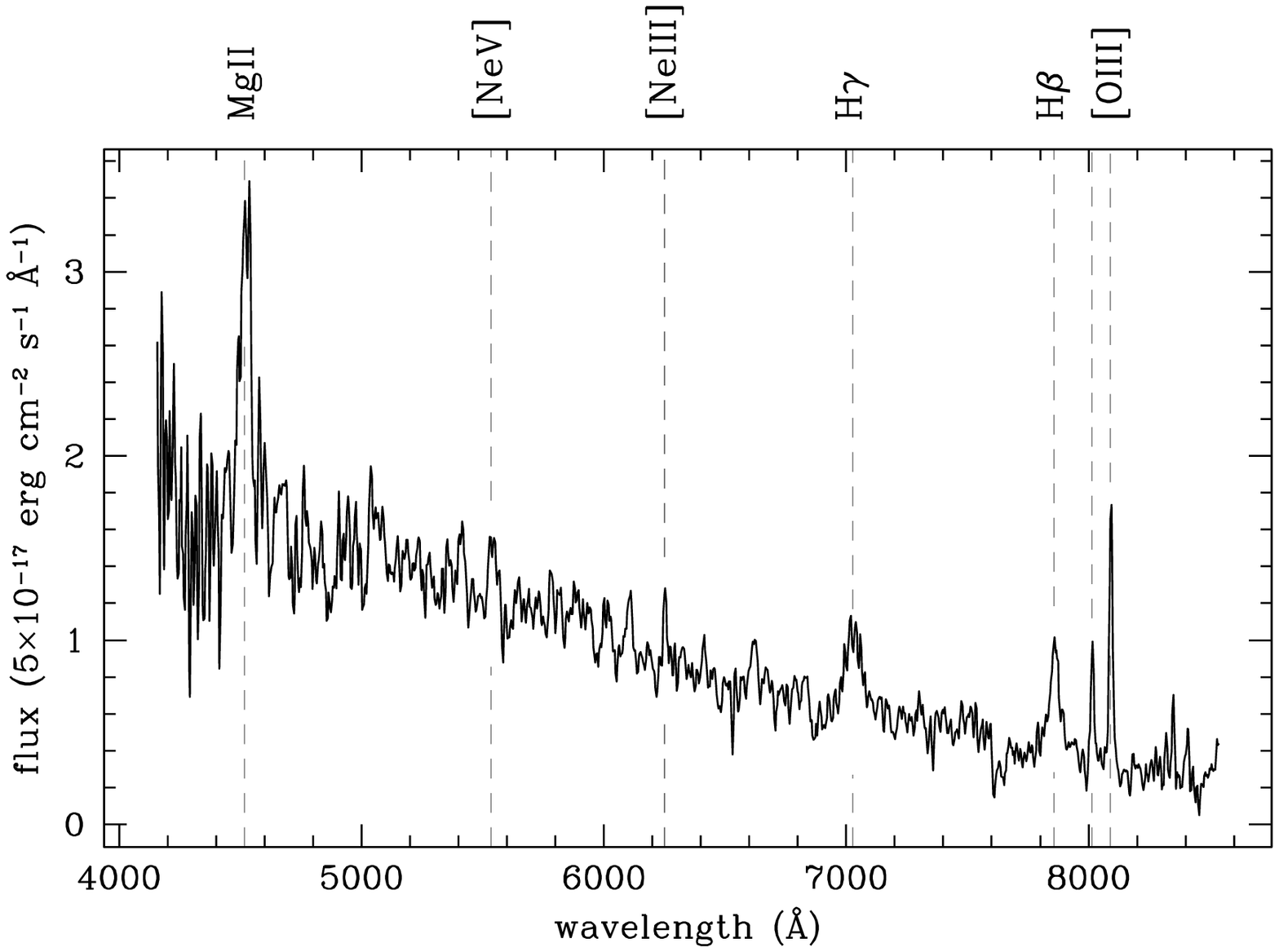}
\caption[RX\,J1813.6$+$6731 spectrum]
{RX\,J1813.6$+$6731: AGN1 at z$=$0.6168$\pm$0.0004 
($\alpha_{2000}=$~18~13~43.0, $\delta_{2000}=+$67~32~23). Longslit 
spectrum of a QSO 
obtained with  the U\,2.2m  telescope on  7 June 1990. The total 
integration time was 30 minutes. The  dashed lines indicate the  
positions of the emission lines at the AGN1 redshift.  Wavelengths 
of atmospheric absorption bands are  also indicated.}
\label{4721}
\end{figure}

\begin{figure}[t]
\epsscale{1}
\plotone{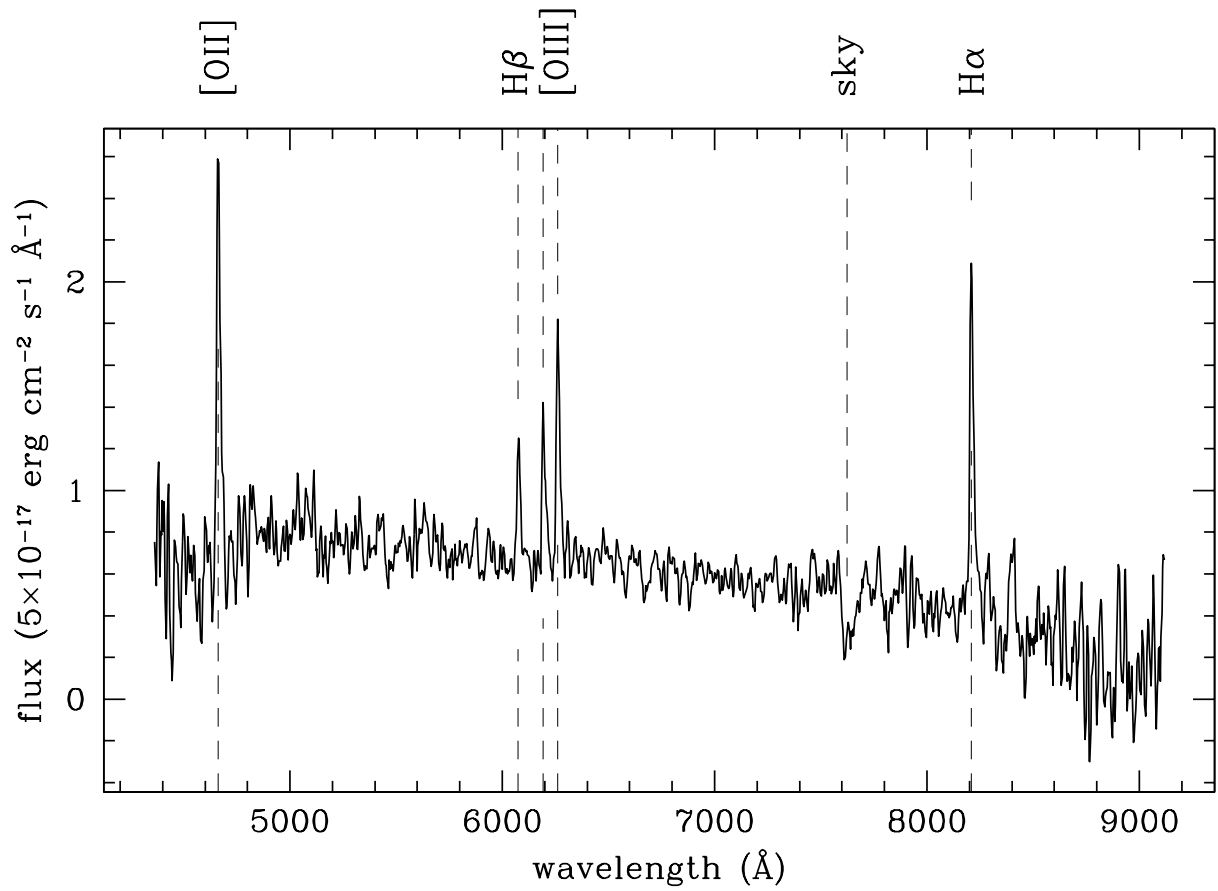}
\caption[RX\,J1750.2$+$6415 spectrum]
{RX\,J1750.2$+$6415: AGN2 at z$=$0.2504$\pm$0.0003 
($\alpha_{2000}=$~17~50~15.1, $\delta_{2000}=+$64~14~56). 
Longslit spectrum of a Sy2 galaxy obtained with the UH\,2.2m 
telescope on 31 July 1998. The total integration time was 20 minutes. The  
dashed lines indicate the  positions of the emission lines at the AGN2 
redshift.  Wavelengths  of atmospheric absorption bands are  
also indicated.}
\label{3060}
\end{figure}

\begin{figure}[t]
\epsscale{1}
\plotone{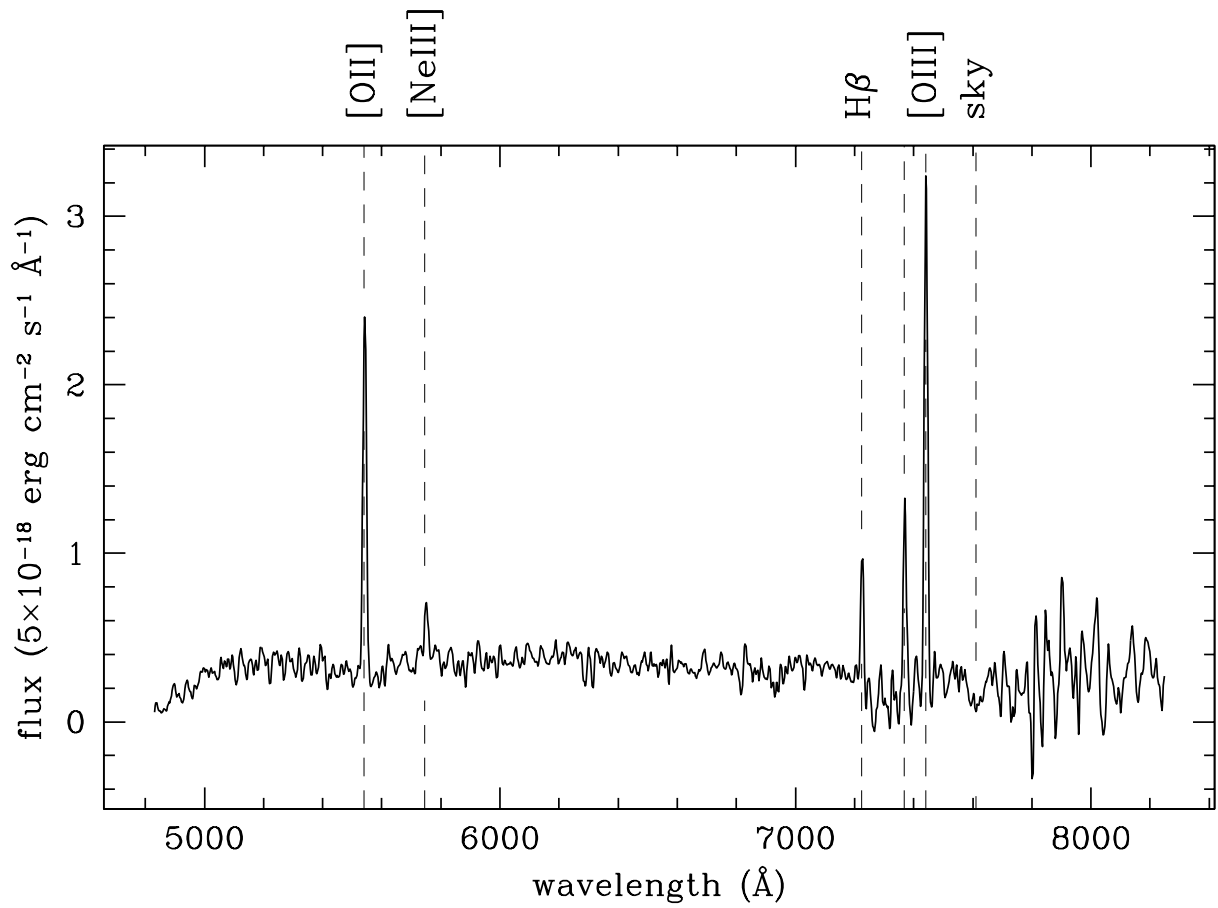}
\caption[RX\,J1757.9$+$6609 spectrum]
{RX\,J1757.9$+$6609: AGN2 at z$=$0.4865$\pm$0.0002 
($\alpha_{2000}=$~17~57~56.5, $\delta_{2000}=+$66~09~20). Longslit spectrum
of a Sy2 galaxy obtained with Keck-II  LRIS on 16
July 1998. The total integration time was 30 
minutes. The dashed lines indicate the  positions of the emission lines at 
the AGN2 redshift.  Wavelengths  of atmospheric absorption bands are  
also indicated.}
\label{210}
\end{figure}

\begin{figure}[t]
\epsscale{1}
\plotone{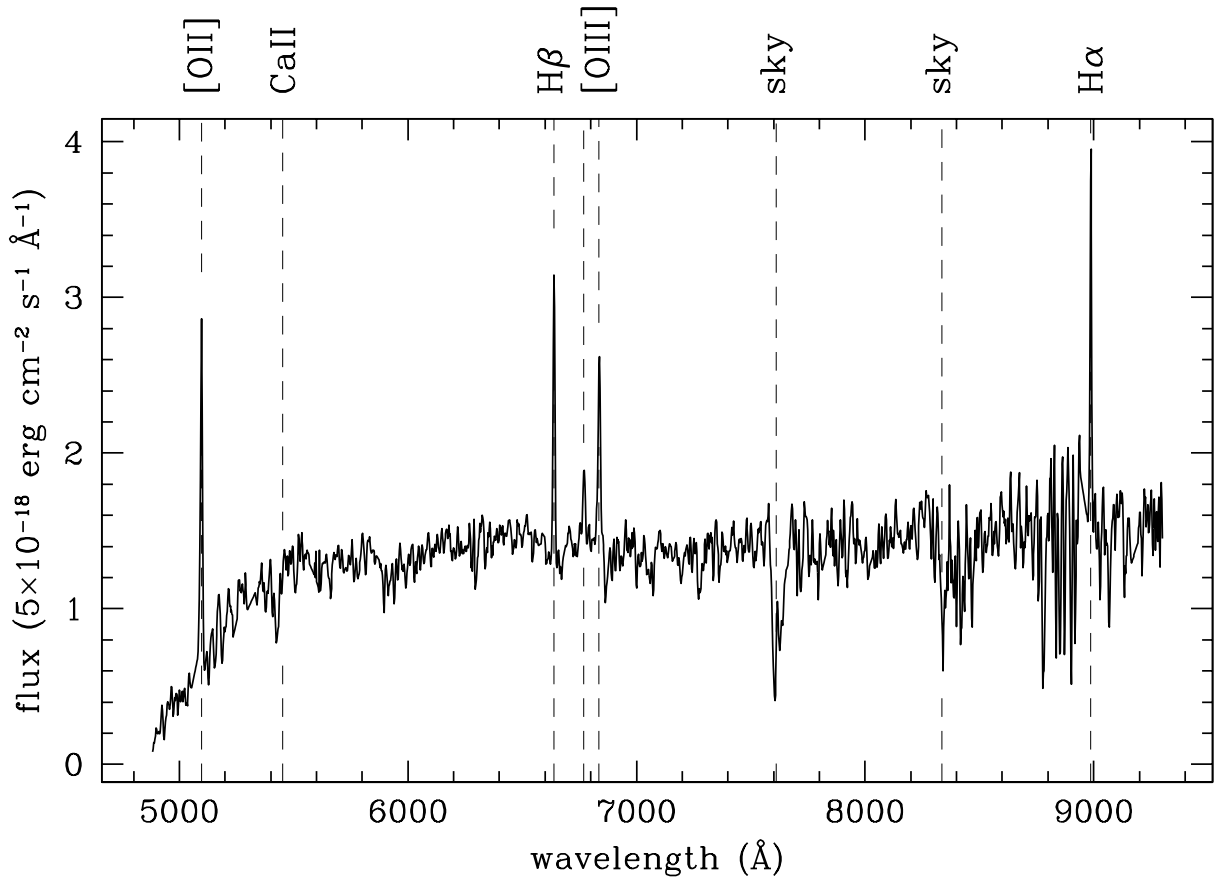}
\caption[RX\,J1758.9$+$6520 spectrum]
{RX\,J1758.9$+$6520: cluster of galaxies at z$=$0.3652$\pm$0.0008. 
Longslit  spectrum of a cluster member, galaxy C ($\alpha_{2000}=$~17~58~53.8,
$\delta_{2000}=+$65~21~02, z$=$0.3665$\pm$0.0008) which shows AGN2 
spectral features. The spectrum was obtained with  Keck-II LRIS
on 16 July 1998. The total integration time was 30 minutes. The  
dashed lines indicate the  positions of emission lines and of the stellar 
absorption features at the AGN2 redshift.  Wavelengths of atmospheric 
absorption bands are also indicated.}
\label{310}
\end{figure}

\begin{figure}[t]
\epsscale{0.9}
\plotone{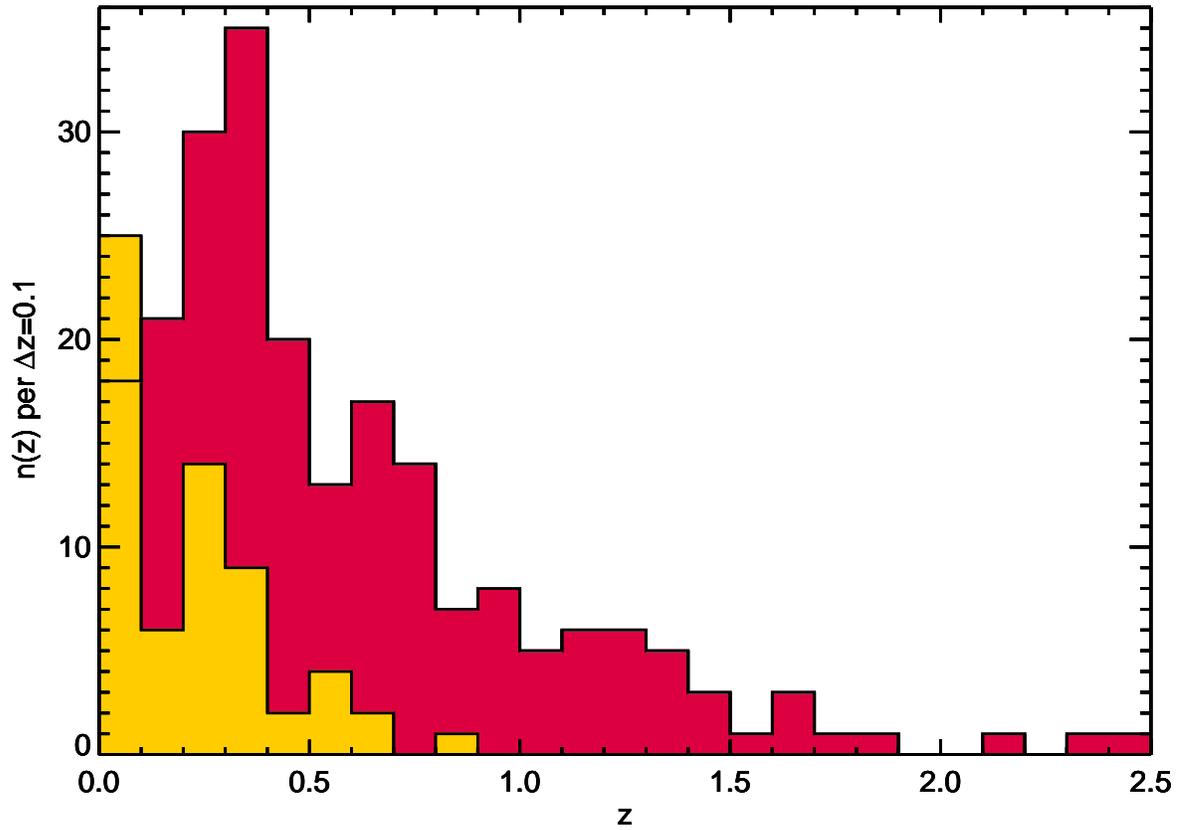}
\caption[Redshift Histogram]
{Redshift distributions of NEP AGN (dark grey) and clusters of galaxies 
(light grey) in bins of $\Delta$z~$=$~0.1. To preserve the  readibility 
of this plot the highest redshift AGN1 (RX\,J1746.2$+$6227,
z$=$3.889, L$_{X}=4.65\times10^{46}$ erg s$^{-1}$) is not shown.}
\label{histo}
\end{figure}

For the classification of the optical counterparts to the NEP
X-ray sources we have followed the pioneering work of \citet{sto91} for the
EMSS  survey. Table~\ref{table2} presents a summary of the 
identifications of the NEP X-ray sources. 
Sixty-five per cent of the extragalactic counterparts
and  nearly half (49.4\%) of all sources are AGN, whose spectra normally are 
characterized by the presence of emission lines and have a blue relatively
featureless continuum. The classification with AGN has been done on the basis
of equivalent width of the emission lines (W$_{\lambda}$) and 
broadness (FWHM) of the permitted emission lines. AGN showing a QSO like 
spectrum with a  W$_{\lambda} \geq 5$\AA, and broad permitted emission 
lines  (FWHM $\geq$ 2000 km s$^{-1}$), have  been classified as AGN 
type~1 (AGN1 in Table~3). This class includes  QSO objects and 
Seyfert 1 galaxies. Examples of this kind of spectra are shown in 
Fig.~\ref{2810}, Fig.~\ref{3620} and Fig.~\ref{4721}. AGN type~2 
(AGN2 in Table~3) have similar permitted and forbidden emission  
lines, both narrow and with a FWHM $<$2000 km s$^{-1}$, mostly  much 
narrower than the limit of  2000 km s$^{-1}$. AGN2 include Seyfert 2 
and star forming galaxies. Examples of AGN2 spectra are shown in 
Fig.~\ref{3060} and   Fig.~\ref{210}. 
Some of the cluster galaxies  in the NEP catalog  show 
spectral similarities to AGN type~2. Even in these cases the
X-ray source is identified as a cluster.  See Fig.~\ref{310} for the 
spectrum of a  Sy2 galaxy at z$=$0.3665 in a cluster at z$=$0.3652. 
There are four spirals with lines having FWHM $<$2000 km s$^{-1}$,
they have been classified as AGN2 and are indicated in Table 3.
This number is however a lower limit since only for the nearby objects
it is possible to make a morphological classification.
The redshift distribution for the AGN in bins of $\Delta$z~$=$~0.1
is shown in Fig.~\ref{histo} (dark grey). The median and highest redshifts 
for AGN are z$=$0.4  and z$=$3.889 (RX\,J1746.2$+$6227), respectively. 

The second most common identification class is Galactic stars at 34.3\%. 
Stellar counterparts to X-ray sources at the Galactic latitude of the 
NEP are usually late-type (F--M) stars whose spectra display hydrogen,
metallic, and molecular absorption features. 
One planetary nebula (NGC 6543) was also detected as an 
X-ray source \citep[RX\,J1758.5$+$6637,][]{kre92}.

\begin{figure}[t]
\epsscale{1}
\plotone{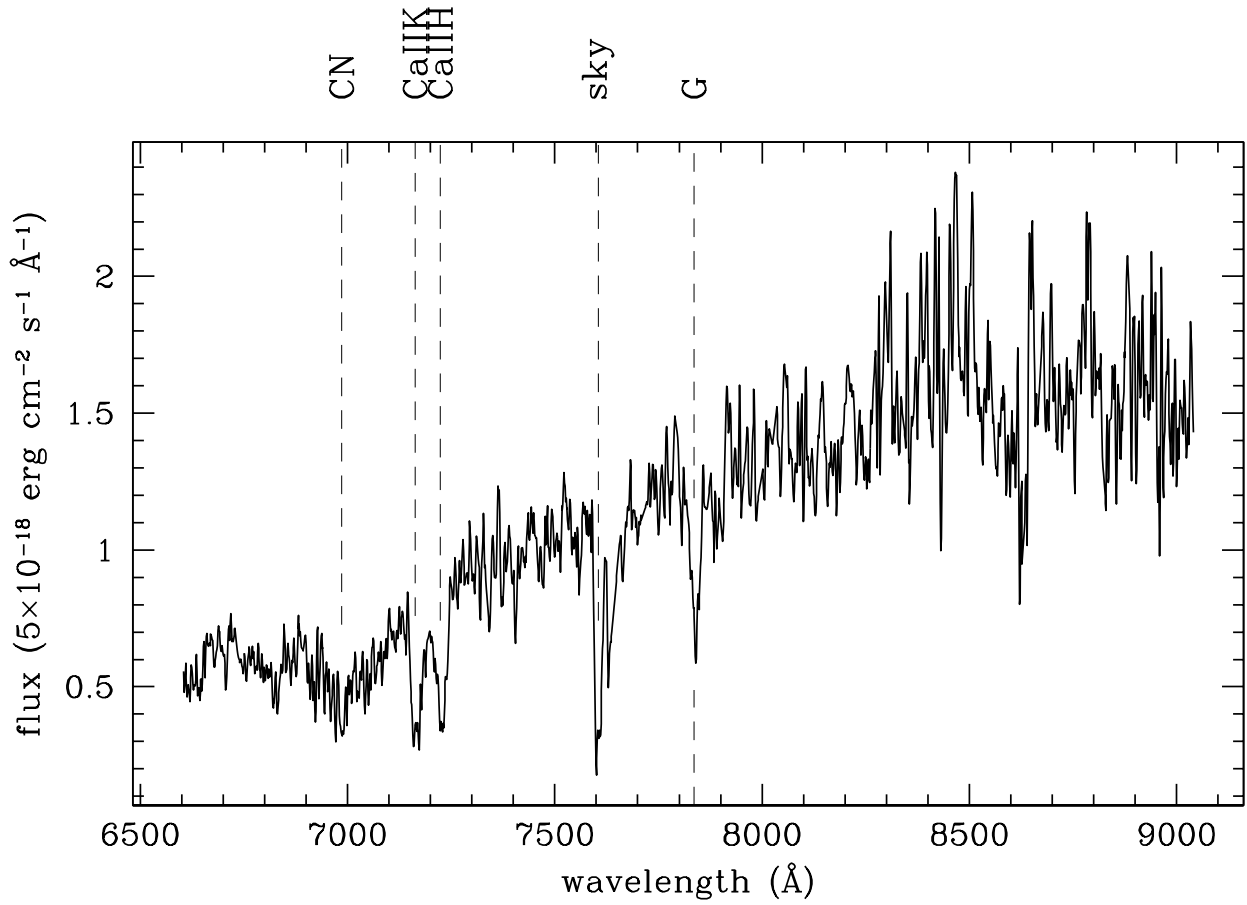}
\caption[RX\,J1821.6$+$6827 spectrum]
{RX\,J1821.6$+$6827: cluster of galaxies at z$=$0.8108$\pm$0.0012. Spectrum
of the cluster galaxy \#116 ($\alpha_{2000}=$~18~21~26.3, 
$\delta_{2000}=+$68~29~28, z$=$0.8202$\pm$0.0001)
obtained with Keck-I LRIS in multislit  mode on 23 June 2001. 
The total integration time was 2.25 hours. The dashed lines indicate the  
positions of stellar absorption features at the cluster redshift.  
Wavelengths of atmospheric absorption bands are also indicated.}
\label{5281}
\end{figure}

\begin{figure}[t]
\epsscale{1}
\plotone{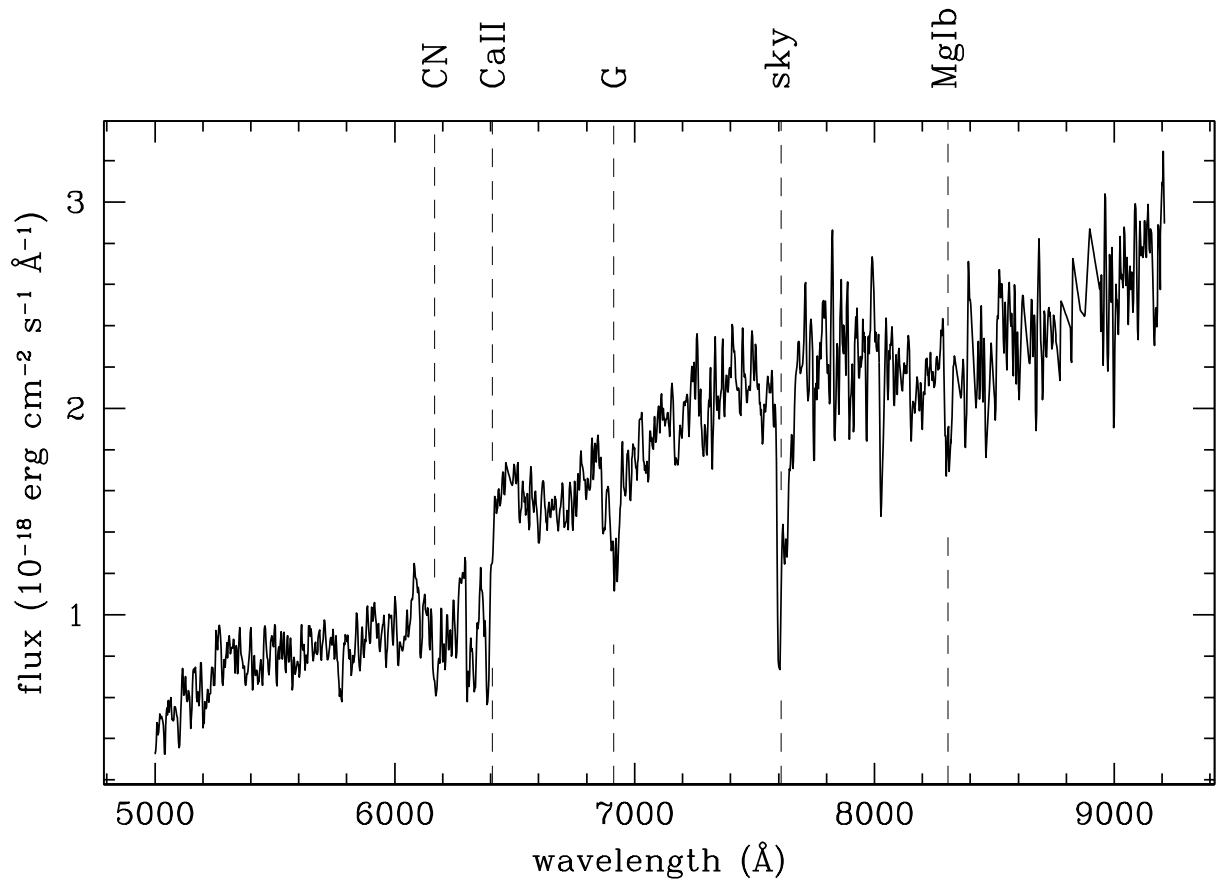}
\caption[RX\,J1745.2$+$6556 spectrum]
{RX\,J1745.2$+$6556: cluster of galaxies at z$=$0.608$\pm$0.0005. 
Longslit spectrum of the cluster galaxy B ($\alpha_{2000}=$~17~45~18.2,
$\delta_{2000}=+$65~55~42, z$=$0.6077$\pm$0.0009) obtained with 
Keck-II LRIS on 26 June 1998. The total integration time was 30 
minutes. The dashed lines indicate the  positions of stellar absorption 
features at the cluster member redshift.  Wavelengths of atmospheric 
absorption bands are also indicated.}
\label{2560}
\end{figure}

\begin{figure}[t]
\epsscale{1}
\plotone{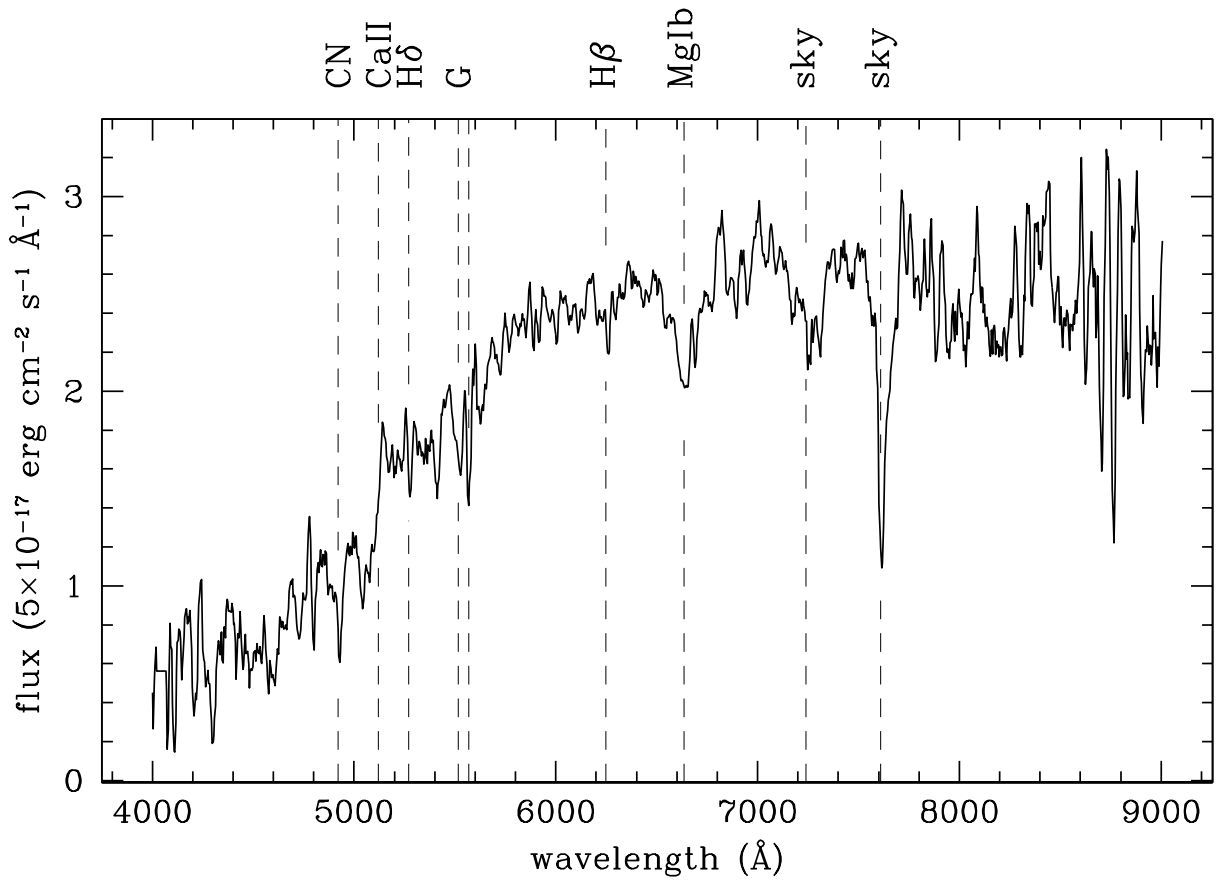}
\caption[RX\,J1817.7$+$6824 spectrum]
{RX\,J1817.7$+$6824: cluster of galaxies at z$=$0.282$\pm$0.002. 
Longslit spectrum of the cluster cD galaxy A ($\alpha_{2000}=$~18~17~44.7,
$\delta_{2000}=+$68~24~24, z$=$0.282$\pm$0.002) obtained with the UH\,2.2m
on 4 August 1994. The total integration time was 30 minutes. The dashed lines 
indicate the  positions of stellar absorption  features at the cluster 
member redshift.  Wavelengths of atmospheric 
absorption bands are also indicated.}
\label{5030}
\end{figure}

Clusters and groups of galaxies comprise 14.0\% of the sources. 
The identification of an X-ray source as a cluster of galaxies 
usually requires the absence of emission lines identifying AGN, 
the absence of a non-thermal continuum identifying a BL Lac object,
coupled with a centrally concentrated galaxy overdensity either from the
POSS or from deep optical CCD images taken for more distant clusters,
and at least two concordant redshifts. Multi-object spectroscopy was  
performed for several fields around X-ray sources which showed  
overdensities of faint galaxies. Many of the sources proved to be 
distant clusters (z$\geq$0.3).  The optical spectrum of a galaxy in  
the highest-z cluster of the NEP survey (RXJ\,1821.6$+$6827,  
z$=$0.811) is shown in Fig.~\ref{5281}. Two more spectra of galaxies in 
clusters (RX\,J1745.2$+$6556 at z$=$0.608 and RX\,J1817.7$+$6824 at 
z$=$0.282) are given in Fig.~\ref{2560} and Fig.~\ref{5030} respectively.
The median and highest  redshift for clusters of galaxies are respectively 
0.205 and 0.811.  Their redshift distribution in bins of $\Delta$z~$=$~0.1
is  shown in Fig.~\ref{histo}  (light grey).

A featureless, blue continuum without any line emission or significant 
4000\AA ~break suggests a BL Lacertae type object. The optical spectrum is 
characterized by  W$_{\lambda} \leq  5$\AA, ~either for emission or
absorption lines, and a  Ca\,II H \& K ``break'' 
(when present)  less than 25\%. The very weak equivalent width limit for any 
emission lines present separates the BL Lac objects from the weak-line AGN.
The limit of 25\% of the flux depression  discontinuity  blueward across the 
Ca\,II  break is significantly less than the 50\% contrast possessed by a 
normal elliptical galaxy. This limit is imposed to ensure the presence of a 
substantial non-thermal continuum and thus discriminate BL Lacs from normal 
galaxies. See extensive discussion of this point in \S\,3.1.2 of
\citet{sto91}. There are 8 BL Lacertae objects in the NEP survey (1.8\%),
of which 4 (RX\,J1727.0+6926, RX\,J1742.7+6852, RX\,J1759.8+7037, 
RX\,J1803.9+6548) are new discoveries.

One source is identified as individual galaxy  (GAL in Table 3). 
This source is associated with an individual, non-AGN galaxy 
(RX\,J1806.4$+$7028, z$=$0.0971). However, this apparently isolated galaxy has 
an X-ray luminosity commensurate with that of a high-luminosity galaxy group 
or low-luminosity galaxy cluster ($\sim$$2 \times 10^{43}$ erg s$^{-1}$, 
0.5 -- 2.0 keV). As first pointed out by Mullis (2001), this is likely 
an example of an X-ray overluminous elliptical galaxy 
\citep[OLEG,][]{vik99} or a ``fossil group'' \citep{pon94}, postulated 
to be the result of galaxy merging within a group.  

There are a few cases where the identification is ambiguous. For instance,
in the field of one source, identified with an AGN1 (RX\,J1720.8$+$6210) 
at z$=$0.7313, there are two galaxies at the same redshift as the AGN1 and 
closer to the X-ray position. In this case the presence of a distant cluster 
with similar redshift is not excluded. Other cases include sources
identified with clusters (i.e. RX\,J1745.2$+$6556 at z$=$0.608, see 
Fig.~\ref{2560} for the spectrum of a member galaxy at z$=$0.6077) 
where  an AGN at a different redshift (z$=$0.2904) from the cluster redshift,
is present in the field and could contribute to the X-ray emission of the 
source.  There are also two  examples of X-ray sources identified as cluster 
plus AGN. Namely, RX\,J1758.9$+$6520 (z$=$0.3652) and RX\,J1806.1$+$6813
(z$=$0.303), are two sources identified as clusters of galaxies even if the 
contribution to  the X-ray emission of an AGN1 (in each case at the same 
redshift as the cluster)  is not excluded.  Other cases include X-rays 
sources with two  AGN present in their error circles.
Only high resolution and high  sensitivity X-ray observations would allow 
us to identify the X-ray source unambiguously. The number of these ambiguous
identifications is very low ($\sim$ 1\%). They are all reported in 
\S\,\ref{Notes} (Notes to Individual Sources). 

Given the higher angular resolution of the {\em ROSAT\/} PSPC ($\sim 25''$) 
compared  to the  Einstein IPC ($\sim 1'$), there was no need to require 
a variety of optical and radio observations in order to conclusively  
identify the class of the optical counterpart to the NEP sources with 
confidence. However,  both space-based and ground-based surveys of the NEP 
region have been performed and the available source catalogs produced by 
those surveys have been scrutinized by us.  We mention here the surveys of 
the NEP region from space completed  by IRAS \citep{hh87}, ISO \citep{sti98, 
aus00} and COBE \citep{ben96}. From the ground, surveys were performed  in 
radio by \citet{kol94, bri99, ren97, loi88, elv94}, and in optical/IR 
by \citet{gai93} and \citet{kw00}.

The distribution in the sky of the 445 NEP X-ray sources and their 
optical identification already appeared in the literature and are shown in 
Fig. 4.12 of \citet{mul01a} and in Fig. 3  of \citet{hen01}.

\section{The Catalog} \label{Catalog}

The 445 sources that form the ROSAT NEP source  catalog  are  presented 
in Table~3. The columns contain the following information:

\begin{enumerate}
\item Source name formed by the acronym
RX\,J $=$ {\bf R}OSAT {\bf X}-ray source, {\bf
J}ulian 2000 position, and the X-ray centroid position. 
\item 
Internal source identification number which runs between 10 and 6570.  
\item 
Right Ascension of the X-ray  centroid (J2000, HH MM SS.S).
\item 
Declination of the X-ray  centroid (J2000, +DD MM SS).
\item 
Right Ascension of the optical object associated with the X-ray source
(J2000, HH MM SS.S).
\item 
Declination of the optical object associated with the X-ray source
(J2000,  +DD MM SS).
\item
Signal-to-Noise on the detected source count rate determined as
net source count rate over 1-$\sigma$ uncertainty on the count rate.
\item
Detect unabsorbed flux in the 0.5--2.0 keV band  ($f_{\rm X,det}$, 
$10^{-14}$ erg cm$^{-2}$ s$^{-1}$).  The detect flux is measured in the 
photometry circular aperture (5\arcmin ~radius). To determine the fluxes 
for the different classes of astronomical objects we have converted from 
count rate to unabsorbed flux using conversion factors based on three 
different types  of source spectra (see \S\,\ref{X-ray}).
\item
Total unabsorbed flux in the 0.5--2.0 keV  band ($f_{\rm X,tot}$, $10^{-14}$ 
erg cm$^{-2}$ s$^{-1}$).  The total flux accounts for the flux
outside the photometry aperture and reflects the size  correction 
applied to the detect flux.  For point sources this flux corretion 
factor is constant and equal to 1.0369 (see \S\,\ref{X-ray}), while it 
varies for extended sources, such as clusters or groups of galaxies. The flux
correction factor for extended sources is given in Column (13).
\item 
Rest frame K--corrected luminosities in the 0.5--2.0 keV 
band ($L_{\rm X}$, $10^{44}$ 
erg s$^{-1}$) for extragalactic objects with uncertainties based 
on the fractional errors on the source count rate. K--correction factors
for clusters of galaxies, assuming a Raymond-Smith plasma spectrum with a 
metallicity 0.3 solar,  are 0.76, 0.95 and 1.01,  minimum, 
median and  maximum values respectively. For AGN the assumed power law 
spectrum with  energy index $= -$1 gives a formal K--correction factor of 1.
\item
Spectroscopically measured redshift for all extragalactic sources. Typical
uncertainty is $\leq$0.001.
\item
Optical identification of the X-ray source: AGN for active galaxtic 
nucleus, either type~1 (AGN1) or type~2 (AGN2) (see \S\,\ref{Idstat}); 
STAR for star;  
CL for group or cluster of galaxies; BL for BL Lacertae object; GAL 
for normal galaxy and PN for Planetary Nebula. Spectral type for stars 
is also indicated if known.
\item
Comments regarding the source, such as size correction factor 
(sc$=${\rm size}$_{corr}$) for galaxy
clusters,  or indication of a Note (n) to the source given in 
\S\,\ref{Notes}.
\end{enumerate}

\section{Notes to Individual Sources} \label{Notes}
  
{\bf RX\,J1716.2$+$6836}: also called RX\,J1716.0$+$6836 in \citet{bol97}. The
identification comes from the revised Burbidge catalog \citep{hew93}.

{\bf RX\,J1717.7$+$6431}: there is a star within the error circle which 
is closer to the X-ray position (7.7\arcsec) than the AGN2. We still believe 
that the AGN2 is the correct identification, not only because it is an AGN
but also because  it appears to be in a distorted spiral in our UH\,2.2m B-band 
image. Distortion implies some kind of interaction and often these distorted
galaxies are X-ray sources.

{\bf RX\,J1719.8$+$6457}: this source is one of two sources in the NEP still 
unidentified. Optical spectra were taken for seven objects within 
80\arcsec from the X-ray position,  but no plausible identification was 
found.

{\bf RX\,J1720.8$+$6210}: there are two galaxies   at the same redshift as 
the AGN1 (z$=$0.7313) and closer to the X-ray centroid (approximate 
positions from DSS-2 red are for cfh\#11, 
$\alpha_{2000}=$~17~20~46.8, $\delta_{2000}=+$62~10~25;
and for cfh\#12b, $\alpha_{2000}=$~17~20~43.9, $\delta_{2000}=+$62~10~11). 
We still identify  the source as AGN1 since its spectrum shows a broad 
MgII emission line  (FWHM$>$4000 km s$^{-1}$). However the presence of a 
distant cluster at z$\sim$0.73 is not excluded. 

{\bf RX\,J1723.1$+$6826}: this source is identified with a QSO 
($\alpha_{2000}=$~17~23~09.9,  $\delta_{2000}=+$68~26~56, z$=$0.9782$\pm$0.001)
which is 4\arcsec away from the X-ray centroid. The optical spcetrum
shows a very broad MgII emission  line (FWHM$\geq$7000 km s$^{-1}$).  A
second object  at $\alpha_{2000}$ $=$~17~23~10.1, $\delta_{2000}=+$68~26~51,
just south  of the QSO  and only 3.4\arcsec away from  the X-ray centroid,  
has a similar  redshift  (z$=$0.9777) as the QSO.
This second object is fainter and has a narrower MgII line in emission. 
Both objects could contribute to the X-ray emission even if we indicate
the object with broader MgII line  as the identification in the table.

{\bf RX\,J1724.1$+$7000} and {\bf RX\,J1724.2$+$6956}: these two sources
were identified as a group of galaxies in \cite{hen95}. The X-ray morphology 
is complex and elongated along the north-south direction (see Fig. 1a in
Henry et al. 1995). We confirm here the identification of these sources  
with a single group of galaxies at z$=$0.0386. 

{\bf RX\,J1727.0$+$6926}:  a radio source (0.96$\pm$0.06 mJy at 6cm)
was detected with the VLA in the DnC array at the position of the optical 
counterpart  by T. Rector (private communication).

{\bf RX\,J1727.8$+$6748}: there are several galaxies in the area for
which no optical spectrum is available. The suggested identification
is the AGN1 at z$=$0.4950 ($\alpha_{2000}=$~17~27~45.5, 
$\delta_{2000}=+$67~48~43) lying 22\arcsec away from the X-ray centroid, 
with a broad (FWHM$>$7000 km s$^{-1}$) MgII emission line in its optical
spectrum.
 
{\bf RX\,J1732.5$+$7031}: this object appears in  the sample of identified 
northern  {\em ROSAT\/} sources by \cite{app98} with a redshift z$=$0.209 vs.
our z$=$0.2114.

{\bf RX\,J1732.9$+$6533}: the redshift for this QSO (z$=$0.8560) comes from 
\cite{hew93}.

{\bf RX\,J1736.3$+$6802}: this group of galaxies was published in 
\cite{hen95}.  There are twelve galaxies with spectroscopic redshifts. 
Refer to  \cite{hen95} for more details and for an X-ray contour image 
(their Fig. 1c).

{\bf RX\,J1736.9$+$6845}: this X-ray source is MS\,1737.2$+$6847
and its identification with SAO17576 ($\Omega$ Draconis) comes from 
the EMSS; see \cite{sto91} and \cite{mac94}. 

{\bf RX\,J1736.4$+$6828}: the X-ray source is identified with GAT 732,
a star with an E magnitude in APM  of 8.98. The star has a high
proper motion (about 1 arcmin in 50 years).

{\bf RX\,J1739.7$+$6710}: this source  is MS\,1739.8$+$6712, and is
identified as AGN1. The  redshift, z$=$0.118, comes from \cite{sto91}; 
an optical finding chart is published  in \cite{mac94}.

{\bf RX\,J1741.2$+$6507}: there is a cluster candidate in the field
of this source identified as AGN1 at z$=$0.7466. We have concordant 
redshifts for two galaxies (A and B): $\alpha_{2000}=$~17~41~15.5, 
$\delta_{2000}=+$65~07~53, z$_{A}=$0.3797; $\alpha_{2000}=$~17~41~07.7, 
$\delta_{2000}=+$65~07~47, z$_{B}=$0.3775.

{\bf RX\,J1743.4$+$6341}: this cluster, associated with A2280, has a large 
gravitational arc which is described in \cite{gio95}.

{\bf RX\,J1745.2$+$6556}: in the field of this source, identified as a cluster 
of galaxies at z$=$0.6080$\pm$0.0005 (optical spectrum for a member galaxy 
is shown in Fig.~\ref{2560}) there is also an AGN 
($\alpha_{2000}=$~17~45~17.6, $\delta_{2000}=+$65~56~02, z$=$0.2904) 
which is $\sim$18\arcsec away from 
the X-ray centroid.  The AGN could contribute to the 
X-ray emission of the source. However, it is difficult to classify the AGN 
as type 1 or type 2 since there is no H$\beta$ emission line, and the 
H$\alpha$ emission line is blended with NII. The spectrum of the AGN has
a very red continuum.

{\bf RX\,J1746.1$+$6737}: the identification of this X-ray source 
comes from the EMSS. This is MS\,1746.2$+$6738 identified as AGN1 at 
z$=$0.041 by \cite{sto91}. The AGN1 ($\alpha_{2000}=$~17~46~08.8,
$\delta_{2000}=+$67~37~15) is 25\arcsec south of a bright SAO star 
(SAO17632, $\alpha_{2000}=$~17~46~08.7, $\delta_{2000}=+$67~37~43, 
m$_{V}=$7.79) which could also contribute to the X-ray emission.

{\bf RX\,J1746.2$+$6227}: the redshift for this QSO (z$=$3.889) is taken from 
Hook et al. (1995; see  their Fig. 2 for a spectrum). \cite{sti93} 
gives a redshift of z$=$3.886 for this object. The QSO was 
independently   discovered as an X-ray source  by \cite{bhw92} who
measured a redshift of z$=$3.87  (optical spectrum in their Fig. 2). 
An X-ray spectrum with ASCA is published in  Fig. 1 of \cite{kub97}.

{\bf RX\,J1747.0$+$6836}: this very bright source is MS\,1747.2$+$6837.  
The identification as AGN1  (z$=$0.063) comes from \cite{kc82} and 
\cite{sto91}. P. F. Winkler  (private communication) originally discovered 
this object.

{\bf RX\,J1747.4$+$6626}: there are two objects here, both AGN1 and 
at the same redshift (z$=$0.1391), which could both contribute to the X-ray 
emission. The optical position for one of them  is 
$\alpha_{2000}=$~17~27~47.0, $\delta_{2000}=+$66~26~24 (given in 
Table 3). The second object, which is a spiral galaxy at z$=$0.1390, is at 
$\alpha_{2000}=$~17~47~26.6, $\delta_{2000}=+$66~26~05.

{\bf RX\,J1747.9$+$6623}: we identify this source as a normal galaxy (GAL) 
but there are two  objects possibly interacting. Object A 
($\alpha_{2000}=$~17~47~58.5, $\delta_{2000}=+$66~23~26, z$=$0.1739) is a 
narrow emission line galaxy with H$\alpha$, 
H$\beta$, H$\gamma$, O[III] emission lines in the optical spectrum, and 
shows the  morphology of a disturbed spiral. Object B 
($\alpha_{2000}=$~17~47~56.8, $\delta_{2000}=+$66~23~46, z$=$0.1737) 
does not show  any emission line in the spectrum and resembles an edge-on 
spiral in our UH\,2.2m  open image.

{\bf RX\,J1748.5$+$7005}: identification and redshift for this BL Lac comes 
from the literature. An optical spectrum is published in Fig. 7 of
\cite{law96}, see also note in \cite{rect01}.

{\bf RX\,J1748.6$+$6842}: there are two AGNs with similar redshift (z$=$0.0537 
and z$=$0.0540) which are blended in the APM finder. The western object 
($\alpha_{2000}=$~17~48~38.3, $\delta_{2000}=+$68~42~17) has
broad emission lines in its optical spectrum. We indicate this object
as the identification (AGN1). The eastern object
($\alpha_{2000}=$~17~48~38.7, $\delta_{2000}=+$68~42~16) has H$\alpha$ in 
emission   but it is difficult to assess the width of the
emission line since the spectrum is rather noisy. Both objects look 
like spiral galaxies in our UH\,2.2m open image. The two objects might be 
interacting and both could contribute to the X-ray emission.

{\bf RX\,J1751.2$+$6533}: this source is associated with the group of galaxies 
published in Fig.~1e of \cite{hen95}. The redshift has been updated using more
accurate low redshift data from \cite{fal99}.

{\bf RX\,J1751.5$+$7013}: this source is identified with a cluster of 
galaxies at z$=$0.4925. The two galaxies, for which we have optical
spectra, have  narrow emission lines typical of AGN2. One of the AGN2 
($\alpha_{2000}=$~17~52~33.1, $\delta_{2000}=+$70~13~01, z$=$0.4936) 
has a narrow MgII line in emission, in addition to H$\beta$ and
to  [OII] and [OIII] lines.

{\bf RX\,J1752.2$+$6522}: this source is identified as a cluster of
galaxies at z$=$0.3923. There is also an emission line object 
($\alpha_{2000}=$~17~52~12.9, $\delta_{2000}=+$65~22~36) which is
15.1\arcsec away from the  X-ray position with z$=$0.3940.
We still identify the source with a cluster of galaxies since
the spectrum of the AGN, with H$\beta$ in emission, is too noisy to assess 
the width of the line.

{\bf RX\,J1753.9$+$7016}: this source is MS\,1754.5+7017, and  is identified
as an AGN1 at z$=$0.062 by \cite{sto91}.

{\bf RX\,J1754.0$+$6452}: the redshift for this cluster is tentative because 
it is based on three low S/N spectra. 

{\bf RX\,J1754.6$+$6803}: this is MS\,1754.9$+$6803. The redshift (z$=$0.0770)
was measured by \cite{sto91} and is based on three galaxy spectra taken at 
the MMT in April  1985.

{\bf RX\,J1756.2$+$7042}: this is one of the two still unidentified sources 
in  the whole NEP survey. The X-ray source is a double source elongated in the
East-West direction. The eastern X-ray centroid is identified with an AGN1  
($\alpha_{2000}=$~17~56~14.9, $\delta_{2000}=+$70~41~56, z$=$0.838) since it 
shows  a broad  MgII line in emission in its optical spectrum. 
The western source is still unidentified. We have taken 
spectra for several objects in the area using the Keck-II
but none of the objects seems to be a satisfactory identification.

{\bf RX\,J1757.2$+$7033}: this source is MS\,1757.7$+$7034, identified by 
\cite{sto91} as a BL Lac at z$=$0.407, from CaII H \& K, G band and MgIb
absorption lines.

{\bf RX\,J1757.2$+$6547}: this X-ray source, identified with an M star
($\alpha_{2000}=$~17~57~14.3, $\delta_{2000}=+$65~46~58, m$_{E}=$14.68),
is 9.4\arcsec away from the X-ray centroid in the APM finders. There is also 
an AGN1 ($\alpha_{2000}=$~17~57~13.6,  $\delta_{2000}=+$65~46~45, z$=$0.578)
15.7 \arcsec away from the X-ray  centroid with m$_{E}=$19.53 and 
m$_{O}=$20.0. The AGN1 could also contribute to the X-ray emission of the 
source.

{\bf RX\,J1757.3$+$6631}: there are five spectroscopic redshifts for this 
cluster at z$=$0.6909. The spectra were taken at the CFHT. None 
of the five galaxies  are visible on the APM, thus the optical positions are 
approximate coordinates measured from  the DSS-2 red plate. Two of the cluster 
galaxies (cfh\#2 at $\alpha_{2000}=$~17~57~46.3,  
$\delta_{2000}=+$66~30~26, z$=$0.7006; and cfh\#8 at 
$\alpha_{2000}=$~17~57~29.9,  $\delta_{2000}=+$66~32~29,  
z$=$0.6860)  are  identified as AGN type~2 galaxies since their 
optical spectra show similar narrow forbidden [OIII], [OII] and permitted 
H$\beta$ and H$\gamma$ emission lines. 

{\bf RX\,J1757.9$+$6609}: this source is identified with an AGN2 
($\alpha_{2000}=$~17~57~56.5,  $\delta_{2000}=+$66~09~20, z$=$0.4865) 
only 4\arcsec away from the X-ray position. See Fig.~\ref{210}
for an optical spectrum of the AGN2. However, spectroscopic 
redshifts for two galaxies ($\alpha_{2000}=$~17~58~02.1,  
$\delta_{2000}=+$66~09~34, and $\alpha_{2000}=$~17~58~00.6, 
$\delta_{2000}=+$66~07~35) which are located 36\arcsec  and 
106\arcsec away from the  X-ray centroid, respectively,
are concordant with the redshift of NEP super cluster (z$=$0.089) 
found in the NEP survey by \citet{mul01b}.

{\bf RX\,J1758.9$+$6520}: this source is identified with a cluster of 
galaxies at z$=$0.3652. Three cluster galaxies spectroscopically 
observed  have narrow emission lines (see Fig.~\ref{310} for an example 
of one cluster member, namely galaxy C at $\alpha_{2000}=$~17~58~53.8, 
$\delta_{2000}=+$65~21~02, z$_{C}=$0.3665). A fourth object in 
the field at $\alpha_{2000}=$~17~59~02.8, $\delta_{2000}=+$65~20~55,  
has broad  emission lines and is thus identified as AGN1 (z$=$0.3660).
The identification of the source is still a cluster of galaxies based on
the distance of the AGN1 from the X-ray position ($\sim35''$), but
a contribution to the X-ray emission from the AGN1 is not excluded.
Thus this  source is a case of cluster plus AGN.

{\bf RX\,J1759.7$+$6629}: this X-ray source (AGN1 at z$=$0.399) was 
already identified by \cite{bow96}. We took an additional spectrum to 
confirm the QSO nature.

{\bf RX\,J1800.0$+$6645}: this X-ray source is the same as  RX\,J1800.0$+$6646
in \cite{bow96} and  was identified  as a G type star. Our spectrum confirms 
that the object is a G-K type star.

{\bf RX\,J1800.1$+$6636}: this X-ray source was already identified by  
\cite{bow96} as a Sy2 galaxy (NGC 6552). We took an additional spectrum  
(z$=$0.0260) to confirm the AGN2 nature.

{\bf RX\,J1801.2$+$6902}: the redshift for this source (AGN1, z$=$1.27)
comes from \cite{lac93}.

{\bf RX\,J1801.2$+$6624}: this X-ray source was already identified in  
\cite{bow96} as a QSO at z$=$1.25.

{\bf RX\,J1801.7$+$6638}: this is a bright X-ray source identified as 
a BL Lac in \cite{bow96} (see finding chart in their Fig. 2-c and an optical
spectrum in their Fig. 6). The redshift is unknown since the optical spectrum 
is featureless. A radio source (1.07$\pm$0.04 mJy at 6cm) was detected with 
the VLA in the DnC array at the position of the optical source by T. Rector
(private communication).

\begin{figure}[t]
\epsscale{1}
\plotone{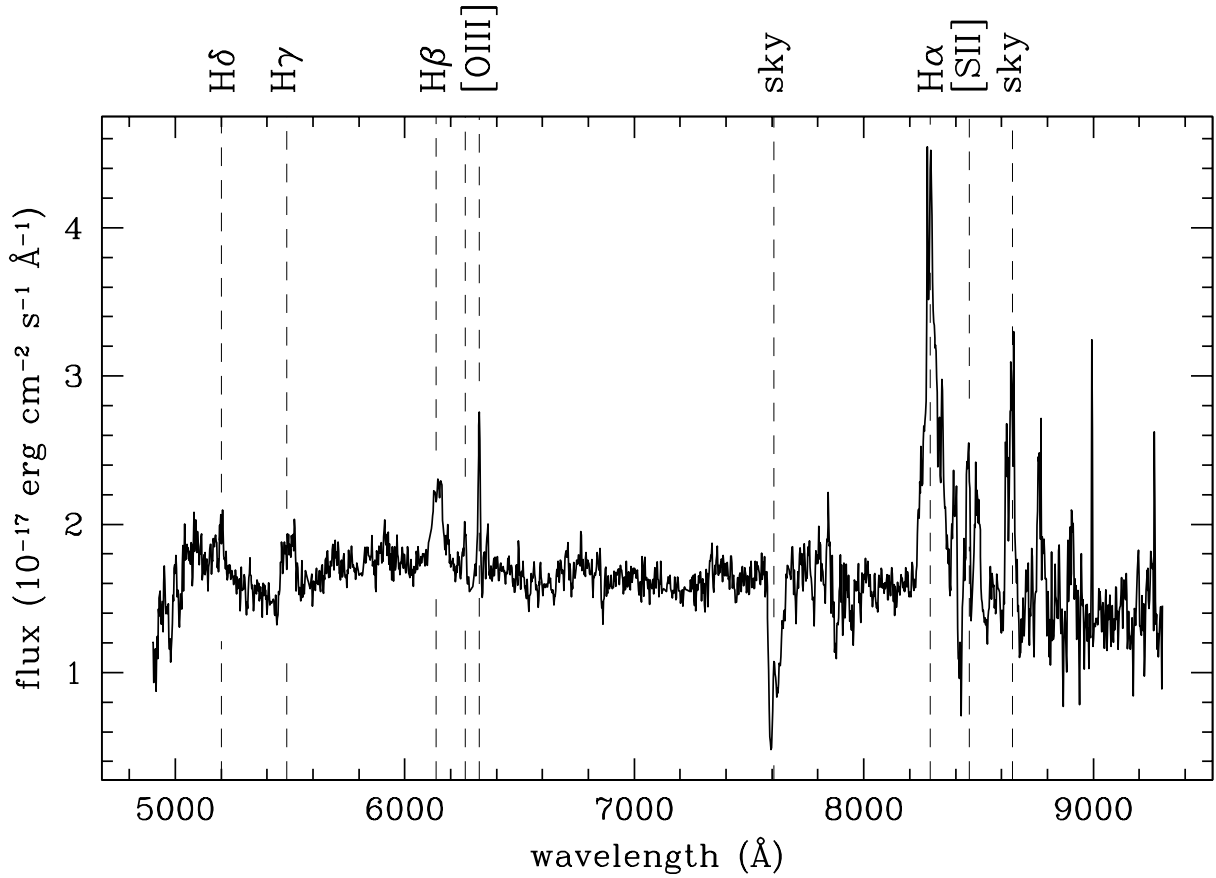}
\caption[RX\,J1802.0$+$6629 spectrum]
{RX\,J1802.0$+$6629: AGN1 at z$=$0.2650$\pm$0.001. 
Longslit spectrum of the  AGN1 ($\alpha_{2000}=$~18~02~04.8, 
$\delta_{2000}=+$66~29~14) obtained with Keck-II LRIS on 21 July 1999. 
The total integration time was 30 minutes. The dashed lines indicate the  
positions of the emission lines at the redshift of the AGN1.  
Wavelengths of atmospheric absorption bands are also indicated. This X-ray 
source was identified as  a BL Lac by \cite{bow96}. Either the BL Lac was 
in a quiescent state or the object identified by  \cite{bow96} is a 
different object.}
\label{560}
\end{figure}

{\bf RX\,J1802.0$+$6629}: this source was identified as a BL Lac in 
\cite{bow96} given the very weak or almost absent lines in the spectrum 
they obtained at the Multiple Mirror Telescope (see their Fig. 3). An optical 
spectrum of this same source was obtained by us  in July 1999 with KeckII-LRIS 
(see Fig.~\ref{560}). Strong and broad  Balmer emission lines are 
visible in our spectrum consistent with an AGN type 1 object, possibly a 
variable QSO?

{\bf RX\,J1802.7$+$6727}: this source is identified as an AGN2 at z$=$0.1620.
There are two possibly interacting spiral galaxies with optical narrow 
emission lines  in their  spectra,  which are blended on the APM finder. 
The optical  position of galaxy A1, at (z$=$0.1605), is 
$\alpha_{2000}=$~18~02~47.8, $\delta_{2000}=+$67~27~41, while the optical 
position of galaxy  A2, at z$=$0.1635, is $\alpha_{2000}=$~18~02~47.6, 
$\delta_{2000}=+$67~27~34. Both galaxies could be responsible for part of 
the X-ray emission.

{\bf RX\,J1803.4$+$6738}: this very bright NEP source is
MS\,1803$+$6728 and is identified as AGN1.  The redshift (z$=$0.1360) was 
measured by \cite{sto91}.  The object is also listed in the 
QSO catalog by  \cite{hew93} as HB89, and in the X-ray NORAS catalog by 
\cite{boe00} as tentative  AGN. Optical spectra taken at 
Lick and Multiple Mirror Telescope observatories are shown in Fig. 4 
of \cite{trev95}.

{\bf RX\,J1803.9$+$6548}: the redshift for this BL Lac (z$=$0.0850)
is tentative since it has  been derived from very weak absorption lines. The 
source is also identified  with the VLA radio source NEP\,J1803.9$+$6548 
by \cite{kol94},  and it is in the radio loud {\em ROSAT\/} NEP sources 
detected with the VLA at 1.5Ghz by \cite{bri99} with a flux$=$43.1$\pm$1.7 
mJy (see more references therein).

{\bf RX\,J1804.3$+$6629}: this source is identified with a very hot,
subdwarf star. Its optical position, $\alpha_{2000}=$~18~04~24.7,
$\delta_{2000}=+$66~29~28, is 26\arcsec away from the X-ray position. 
There is a second  object in the field which is closer 
($\alpha_{2000}=$~18~04~20.6, $\delta_{2000}=+$66~29~54) to the
center of the X-ray emission and for which no spectrum is available. 
However this second object is quite faint, is not blue on the DSS, 
does not have any radio emission, and thus unlikely to be an AGN.

{\bf RX\,J1806.1$+$6813}: this source is identified as cluster of galaxies
at z$=$0.303. Four out of six galaxies for which we have taken 
spectra are emission line galaxies, all with concordant redshifts. One of the
four galaxies ($\alpha_{2000}=$~18~06~04.8, $\delta_{2000}=+$68~13~08, 
z$=$0.2953) has broad Balmer (FWHM$=$2500 kms$^{-1}$). 
Thus this  source is a case of cluster plus AGN.

{\bf RX\,J1806.4$+$7028}:  this source is identified as a galaxy  and
has an X-ray luminosity  L$_{X}=1.8\times10^{43}$ erg ~s$^{-1}$ 
(assuming $kT=$2 keV), which is high for a galaxy with no emission lines. 
As first noted by Mullis (2001) it could be an example of overluminous 
galaxies found in X-ray surveys  (e.g. \citealp{vik99}, and references 
therein).

{\bf RX\,J1806.8$+$6949}: this is a BL Lac in a cluster of galaxies.
Redshift comes from \cite{fal99}.

{\bf RX\,J1808.8$+$6634}: the redshift for this source, identified as
AGN1 at z$=$0.697, comes from \cite{lau98}. See their Table 2 for 
source properties.

{\bf RX\,J1810.3$+$6328}: this source, identified as AGN1 at z$=$0.838,
has a double morphology in X-rays. The identification of the X-ray source
given in Table 3 refers  to the western lobe ($\alpha_{2000}=$~18~10~31.1,
 $\delta_{2000}=+$63~28~08, z$=$0.838). The eastern lobe is an AGN1 at 
z$=$1.0907 ($\alpha_{2000}=$~18~10~16.9,  $\delta_{2000}=+$63~29~14).

{\bf RX\,J1815.4$+$6806}: the  redhsift for this  QSO (z$=$0.239)
comes from \cite{lac93}.

{\bf RX\,J1821.9$+$6420}: the identification of this source as a QSO 
at z$=$0.2970 comes from  Pravdo \& Marshall (1984) (see an optical 
spectrum in their  Fig. 2). However, as already noted by \cite{hn91} 
(see their Fig. 1), a cluster of galaxies for which six galaxy spectra 
were taken by \citet{schn92} is present at the same redshift
as the QSO (see detailed description of this QSO/cluster source in 
\citealp{wol02} and references therein). The X-ray source has the 
highest S/N of the whole NEP survey, it is extremely X-ray luminous
(L$_{X}=5.61\times10^{45}$ erg s$^{-1}$) and it is also an IRAS source, 
as reported by \cite{deg92}.

{\bf RX\,J1826.6$+$6706}: this source is identified with an AGN1 at z$=$0.287.
The redshift comes from \cite{lac93}.

{\bf RX\,J1832.5$+$6848}: this source is identified as a cluster of
galaxies at z$=$0.205. We took optical spectra for three galaxies and 
they have concordant redshifts. The source
is listed  in the literature as the radio source 7C 1832+6845, and it is 
identified as QSO by \cite{vv01},  and as BL Lac in the NORAS catalog 
by \cite{boe00}. The ambiguity may be due to the fact that there are
two distinct objects very closely separated ($\sim$6$''$)
at the position of  the X-ray source. 
The north-west object, object A, is the object we have observed 
spectroscopically ($\alpha_{2000}=$~18~32~35.6, $\delta_{2000}=+$68~48~09,
z$=$0.2049). It is  red on the DSS and its spectrum
shows narrow Balmer and oxygen  emission lines, consistent with an 
AGN2 object (unless we caught the QSO in a quiescent state). The south-east
object A1 ($\alpha_{2000}=$~18~32~36.2, $\delta_{2000}=+$68~48~04)
appears  blue on the DSS and could be the object indicated as QSO or BL Lac 
in the literature. No spectrum is available for A1.
A second galaxy in the cluster (object B at  $\alpha_{2000}=$~18~32~35.9,
$\delta_{2000}=+$68~47~43, z$=$0.2048) also shows narrow emission lines 
and it is  classified as AGN2 in cluster. The third galaxy  
for which we obtained a spectrum, object C ($\alpha_{2000}=$~18~32~35.6, 
$\delta_{2000}=+$68~47~58, z$=$0.2054), shows no emission lines in its
spectrum.

{\bf RX\,J1834.1$+$7057}: this source is very extended in X-rays and it 
is identified with a cluster of galaxies at z$=$0.0803. An extraction radius 
of 6.5\arcmin~ has been used, different from the normal 5\arcmin~ radius 
used for the rest of the X-ray sources.  The X-ray position is centered on a 
bright galaxy ($\alpha_{2000}=$~18~34~08.5, $\delta_{2000}=+$70~57~19,
m$_{E}=\sim$12.5), possibly the cD, whose  redshift appears in the NORAS 
catalog of \cite{boe00} as z$=$0.0824. The source is in the same region of 
sky as A2308, even if the Abell cluster position in the NED database 
is at $\alpha_{2000}=$~18~33~33.8 and $\delta_{2000}=+$71~01~28. 
The redshift given in Table 3 has been computed using  spectra for three
galaxies taken by us at the UH\,2.2m. The redshift agrees with other 
redshift determinations listed in NED for A2308.

{\bf RX\,J1842.5$+$6809}: redshift for this AGN1, z$=$0.4750, comes from 
the literature. The QSO is listed  in the catalog by \cite{hew93} and 
in \cite{xu94}.

\section{Summary} \label{summary}

We have presented data for a survey performed at X-ray wavelengths using
the {\em RASS\/} data in a 80.7 deg$^{2}$ contiguous area of sky at the 
North Ecliptic Pole.  The NEP survey is centered at 
$\alpha_{2000} = 18^{h} 00^{m},~\delta_{2000} =+66\deg 33\arcmin$, and
is at a moderate galactic latitude of {\em b\/} $=$ 29.8\deg.
The NEP catalog consists of a homogeneous, flux-limited sample of 445 
individual X-ray sources above a flux of $\sim$ 2$\times$10$^{-14}$ erg 
cm $^{-2}$ s$^{-1}$ in the 0.5$-$2.0 keV energy band.
The main results of this paper are the optical identifications of the
X-ray sources of the NEP survey. Basic X-ray and optical 
properties of the sources are presented here while finding charts for
all the sources with overlayed X-ray contours will be made available
in a separate publication (see also Appendix C in \citealt{mul01a}).
We  have described in detail the optical identification 
procedure. We have identified 443 out of 445 X-ray source  (99.6\%) and 
determined  spectroscopic redshifts for the extragalactic objects.
All the NEP sources are identified with previously known classes of
X-ray emitters. The optical content of the survey can be summarized as 
follows: 218 AGN (49.4\%), 152 stars (34.3\%), 62 clusters of galaxies 
(14.0\%), 8 BL Lacertae objects (1.8\%),  1 individual galaxy (0.2\%) 
and 1 planetary nebula (0.2\%). 
Given the completeness of the optical identification
and the well defined selection criteria, the NEP survey can
be used to characterize the evolutionary properties of the extragalactic
populations. Evidence  for cluster X-ray luminosity negative evolution 
using the NEP clusters has already appeared in \citet{gio01} while
the X-ray evolutionary properties of the NEP AGN will be the subject
of future publications.

\acknowledgments

IMG notes that this paper was written in spite of the continued 
efforts by the Italian government to dismantle publicly-funded 
fundamental research.
We are grateful to the University of Hawai$'$i Telescope Allocation 
Committee for its generous support of this program. The staffs of 
the University of Hawai$'$i, Canada-France-Hawai$'$i,  Keck,
Multiple Mirror Telescope and F. L. Whipple Observatory telescopes 
performed with their customary expertise. Many thanks are also due to the 
{\em ROSAT\/} team, particularly G\"unther Hasinger and Joachim Tr\"umper. 
We would like to thank Travis Rector at NRAO for observing with the Very
Large Array two NEP sources, whose radio detection confirmed their
BL Lac nature. We are grateful  to the  sponsoring agencies who have 
enabled this project. 
Support has come  from  the NSF (AST 91-19216 and AST 95-00515), NASA 
(NGT5-50175,  GO-5402.01-93A, and GO-05987.02-94A), the ARCS 
Foundation, the Smithsonian   Institution, NATO (CRG91-0415), the Italian 
Space Agency  ASI-CNR,  the  Bundesministerium f\"ur  Forschung 
(BMBF/DLR) and the Max-Planck-Gesellschaft  (MPG).
This research has made use of the NASA/IPAC Extragalactic database (NED)
which is operated by the Jet Propulsion Laboratory, California Institute 
of Technology, under contract with the National Aeronautics and Space 
Administration.

\clearpage
\clearpage
\clearpage
\clearpage
\clearpage
\clearpage
\clearpage
\clearpage
\clearpage
\clearpage
\clearpage
\clearpage
\clearpage
\clearpage
\clearpage
\begin{deluxetable}{llllrrrrr}
\tablecolumns{9}
\tablewidth{5in}
\tablenum{1}
\tablecaption{Optical Follow-up Observations}
\tablehead{
\colhead{Telescope} & & & &\colhead{Nights}&&  & &\colhead{Nights}\\
  &\colhead{} & & &\colhead{Assigned} & &  & & \colhead{Clear}}
\startdata
Mauna Kea &       &             & & 126     &     &  &   & 100  \\
          &       & UH\,2.2m    & & 101     &     &  &   &      \\
          &       & CFHT\,3.6m  & & 17      &     &  &   &      \\
          &       & Keck\,10m   & &  8      &     &  &   &      \\
Mt. Hopkins     & &             & &  6      &     &  &   &   3  \\
          &       & MMT         & &  5      &     &  &   &   2  \\
          &       & 1.5m        & &  1      &     &  &   &   1  \\
TOTAL     &       &             & & 132     &     &  &   & 103  \\
\enddata
\label{table1}
\end{deluxetable}
\clearpage
\begin{deluxetable}{lccccc}
\tablecolumns{6}
\tablewidth{4.5in}
\tablenum{2}
\tablecaption{Summary of the ROSAT NEP \mbox{X-ray} Source Identifications}
\tablehead{\colhead{Type} & & \colhead{Number} & & & \colhead{Percentage\tablenotemark{\dag}}}
\tablecolumns{6}
\tablewidth{4.5in}
\tablenum{2}
\tablecaption{Summary of the ROSAT NEP \mbox{X-ray} Source Identifications}
\tablehead{\colhead{Type} & & \colhead{Number} & & & \colhead{Percentage\tablenotemark{\dag}}}
\startdata
AGN              &\ & 219 & & & 49.4\% \\
star             & & 152 & & & 34.3\% \\
galaxy cluster   & & \phantom{0}62 & & & 14.0\% \\
BL Lac           & & \phantom{00}8 & & & \phantom{0} 1.8\% \\
galaxy           & & \phantom{00}1 & & & \phantom{0} 0.2\% \\
planetary nebula & & \phantom{00}1 & & & \phantom{0} 0.2\% \\
no identification& & \phantom{00}2 & & & \phantom{0} 0.4\% \\
\tablenotetext{\dag}{Out of 443 for identifications, out of 
445 for non-identifications}
\enddata
\label{table2}
\end{deluxetable}
\clearpage
\begin{deluxetable}{cccccccccccll} 
\tablenum{3}
\rotate
\tablewidth{0pt}
\tabletypesize{\scriptsize}
\tablecaption{{\em ROSAT\/} NEP Source Catalog} 
\tablehead{\colhead{Object} & \colhead{NEP}     & \colhead{$\alpha_{\rm X}$}  & \colhead{$\delta_{\rm X}$}  & \colhead{$\alpha_{\rm opt}$}  &  \colhead{$\delta_{\rm opt}$} &  \colhead{S/N}     &  \colhead{$f_{\rm X,det}$} &  \colhead{$f_{\rm X,tot} $} &  \colhead{$L_{\rm X} $} &   \colhead{z}    &  \colhead{ID}   &  \colhead{Notes} \\
\colhead{ }   & \colhead{scan \#} & \colhead{(J2000)} & \colhead{(J2000)}  &  \colhead{(J2000)}  & \colhead{(J2000)} & \colhead{}  & \colhead{$10^{-14}$} & \colhead{$10^{-14}$}  & \colhead{$10^{44}$}  & \colhead{} &  \colhead{} & \colhead{ }  \\
\colhead{(1)}    &  \colhead{(2)}     &  \colhead{(3)}       & \colhead{(4)}       &\colhead{ (5)}   & \colhead{(6)}   & \colhead{(7)}  & \colhead{ (8)}    & \colhead{(9) }  & \colhead{(10)}  & \colhead{(11)}       & \colhead{(12)} & \colhead{(13)}}
\startdata
RX\,J1715.4+6239 & 1239 & 17 15 25.3 & +62 39 34 & 17 15 25.7 & +62 39 27 & \phantom{0}4.6\phantom{0}&\phantom{0}\phantom{0}\phantom{0}18.90\phantom{0}&\phantom{0}\phantom{0}\phantom{0}19.60\phantom{0}&\phantom{0} $ \phantom{0}\phantom{0}8.09\phantom{0} $ & 0.8500 & AGN2 & \\
RX\,J1715.6+6856 & 1240 & 17 15 41.4 & +68 56 31 & 17 15 41.7 & +68 56 43 & \phantom{0}5.3\phantom{0}&\phantom{0}\phantom{0}\phantom{0}\phantom{0}8.81\phantom{0}&\phantom{0}\phantom{0}\phantom{0}\phantom{0}9.14\phantom{0}&\phantom{0}$\cdots$ &$\cdots$ &STAR  & \\
RX\,J1715.6+6231 & 1241 & 17 15 41.6 & +62 31 33 & 17 15 42.3 & +62 31 27 & \phantom{0}4.4\phantom{0}&\phantom{0}\phantom{0}\phantom{0}10.12\phantom{0}&\phantom{0}\phantom{0}\phantom{0}10.49\phantom{0}&\phantom{0}$\cdots$ &$\cdots$ &STAR  & \\
RX\,J1716.2+6836 & 1270 & 17 16 14.4 & +68 36 36 & 17 16 13.8 & +68 36 38 & 20.2\phantom{0}&\phantom{0}\phantom{0}138.48\phantom{0}&\phantom{0}\phantom{0}143.59\phantom{0}&\phantom{0} $ \phantom{0}48.70\phantom{0} $ & 0.7770 & AGN1 & n \\
RX\,J1716.6+6410 & 1271 & 17 16 39.7 & +64 10 35 & 17 16 42.3 & +64 10 41 & \phantom{0}6.5\phantom{0}&\phantom{0}\phantom{0}\phantom{0}23.88\phantom{0}&\phantom{0}\phantom{0}\phantom{0}30.91\phantom{0}&\phantom{0} $ \phantom{0}\phantom{0}0.86\phantom{0} $ & 0.2507 & CL & sc=1.295 \\ 
RX\,J1717.1+6401 & 1272 & 17 17 08.1 & +64 01 46 & 17 17 07.1 & +64 01 45 & \phantom{0}9.3\phantom{0}&\phantom{0}\phantom{0}\phantom{0}44.81\phantom{0}&\phantom{0}\phantom{0}\phantom{0}46.47\phantom{0}&\phantom{0} $ \phantom{0}\phantom{0}0.38\phantom{0} $ & 0.1334 & AGN1 & \\
RX\,J1717.5+6559 & 1300 & 17 17 35.6 & +65 59 35 & 17 17 37.9 & +65 59 39 & \phantom{0}9.0\phantom{0}&\phantom{0}\phantom{0}\phantom{0}53.45\phantom{0}&\phantom{0}\phantom{0}\phantom{0}55.42\phantom{0}&\phantom{0} $ \phantom{0}\phantom{0}2.33\phantom{0} $ & 0.2936 & AGN1 & \\
\enddata
\label{table3}
\end{deluxetable}
\end{document}